\documentclass[journal,draftcls,onecolumn,12pt,twoside]{IEEEtranTCOM}
\normalsize

\ifCLASSINFOpdf
\else
\fi

\usepackage[utf8]{inputenc}
\usepackage{cite}

\usepackage{latexsym,epsfig,amssymb,amsfonts,eucal,graphicx}
\usepackage{subfigure}
\usepackage[cmex10]{amsmath}
\usepackage{slantsc}
\usepackage{enumerate}
\usepackage{setspace}
\usepackage{graphicx}
\usepackage{epstopdf}
\usepackage{mdwmath}
\usepackage{mdwtab}
\usepackage{color}
\usepackage{soul}
\begin{document}
%
\title{Efficient MIMO Transmission of PSK Signals with a Single-Radio Reconfigurable Antenna}
%
%
%

\author{Mohsen~Yousefbeiki,~\IEEEmembership{Student~Member,~IEEE,}
        Osama~N.~Alrabadi,~\IEEEmembership{Member,~IEEE,}
        and~Julien~Perruisseau-Carrier,~\IEEEmembership{Senior~Member,~IEEE}
\thanks{Manuscript received June 26, 2013; revised October 23, 2013 and December 10, 2013. This work was supported by the Swiss National Science Foundation (SNSF) under grant nº133583. }%
\thanks{M. Yousefbeiki and J. Perruisseau-Carrier are with the group for Adaptive MicroNanoWave Systems, Ecole Polytechnique Fédérale de Lausanne (EPFL), CH-1015 Lausanne, Switzerland (e-mails: mohsen.yousefbeiki@epfl.ch, julien.perruisseau-carrier@epfl.ch).}
\thanks{O. N. Alrabadi is with the Antennas, Propagation and Radio Networking (APNet) Section, Department of Electronic Systems, Aalborg University, Denmark (e-mail: osama.rabadi@gmail.com).} 
}

%
%

\markboth{IEEE Transactions on Communications}%
{Post-Print Version}
%



\maketitle

\vspace{-1em}
\begin{abstract}
Crucial developments to the recently introduced signal-space approach for multiplexing multiple data symbols using a single-radio switched antenna are presented. First, we introduce a general framework for expressing the spatial multiplexing relation of the transmit signals only from the antenna scattering parameters and the modulating reactive loading. This not only avoids tedious far-field calculations, but more importantly provides an efficient and practical strategy for spatially multiplexing PSK signals of any modulation order. The proposed approach allows ensuring a constant impedance matching at the input of the driving antenna for all symbol combinations, and as importantly uses only passive reconfigurable loads. This obviates the use of reconfigurable matching networks and active loads, respectively, thereby overcoming stringent limitations of previous single-feed MIMO techniques in terms of complexity, efficiency, and power consumption. The proposed approach is illustrated by the design of a realistic very compact antenna system optimized for multiplexing QPSK signals. The results show that the proposed approach can bring the MIMO benefits to the low-end user terminals at a reduced RF complexity.
\end{abstract}


%
\IEEEpeerreviewmaketitle

\section{Introduction}
%
%
%
%
\IEEEPARstart{D}{espite} its advantageous impact on spectral efficiency, the implementation of the conventional multi-input multi-output (MIMO) concept with multiple RF feeds in small and low-cost communication devices is subject to several design challenges. First of all, the integration of conventional MIMO architectures in small platforms adds complexity and cost constraints since multiple radio frequency (RF) chains are required. Moreover, due to likely spurious emission and imperfect filtering, extreme care should be taken in order to mitigate the self-interference among the parallel RF chains. To overcome this limitation, several MIMO architectures with
reduced RF hardware complexity have been recently proposed in the literature such as antenna selection \cite{antenna-selection}, analogue antenna combining \cite{antenna-combining}, time-division multiplexing \cite{TDM}, code-modulated path-sharing \cite{CMP} and spatial modulation \cite{renzo}. However, most of the aforementioned techniques are applicable only at the receiver side and do not support spatial multiplexing of independent signals with a single RF front-end. Spatial modulation, though being an open-loop transmit technique, achieves a logarithmic increase of the spectral efficiency with the number of transmit antennas compared to the linear growth provided by spatial multiplexing. In fact, spatial multiplexing, unlike the other MIMO modes such as diversity and power focusing (beam-forming), is often referred to as the \textit{true} MIMO mode.

To this end, the authors in \cite{kalis2008} proposed the idea of mapping different symbols onto an orthogonal set of angular basis functions in the beam-space domain of a single-feed switched parasitic array. More precisely, it was assumed that the instantaneous radiation field of the antenna system, $\CMcal{E}_\text{inst}({\theta{,}\varphi})$, can be expressed at any instant of time as a weighted sum of basis functions, $\CMcal{B}_n( {\theta{,}\varphi } )$, such that $\CMcal{E}_\text{inst}( {\theta{,}\varphi } ) = \sum {{s_n}\CMcal{B}_n( {\theta{,}\varphi } )} $ where ${s_n}$ is an arbitrary complex data symbol from the signal constellation diagram. Therefore, in rich-scattering environments, decorrelation between the channel coefficients is guaranteed and thereby the transmitted mixture of information can be reliably decoded using a traditional MIMO receiver.

The second challenge is to maintain high radiation efficiency of the multiple antenna system, which can be compromised by mutual coupling when packing different antenna elements in small platforms. The isolation between a pair of coupled antennas is traditionally achieved by creating an artificial reverse path to the coupling path, using for example a decoupling network \cite{decoupling1,decoupling2}. However, such networks have negative impact on the bandwidth of the multiple antenna system besides being complex and lossy. Remarkably, the efficiency issue is naturally addressed by the novel single-feed MIMO concept. Indeed, unlike in conventional MIMO systems, here mutual coupling is utilized as a controlled signal modulator, and do not entail any power loss.

Another challenge of conventional MIMO lies in maintaining independent and identically distributed (i.i.d.) sub-channels. This is typically addressed by placing the multiple antennas far enough from each other, a requirement that is not practical in real-life user terminals with strict size constraints. In the single-feed MIMO, such sub-channels can be achieved by mapping the data symbols onto an orthonormal set of basis functions. While the orthogonality is guaranteed by the basis definition, the power balance is obtained by optimizing the variable reactive loading of the antenna system and/or the antenna structure as detailed later.

Although the novel single-radio MIMO technique revolutionizes the RF chain in MIMO transmission by reducing the RF hardware size and complexity while staying power efficient, its advantages come with a number of limitations that do not exist in conventional MIMO systems. First, the system is inherently narrowband since the reactive terminations that modulate the data sub-streams are frequency dependent. Moreover, modern modulation schemes such as orthogonal frequency-division multiplexing (OFDM) convert simple constant envelope modulations like PSK into highly complex signal constellation diagrams, which are extremely difficult to emulate with realistic single-radio MIMO hardware. Accordingly, the novel single-radio MIMO approach will find its way to several RF transceiver architectures like the ones used in wireless sensor nodes as well as wireless modems supporting constant envelop signal formats.

Single-radio MIMO has been addressed in several previous publications such as \cite{alrabadi_PSK,barousis,alrabadi_tap,julien_eumc,my_tap,kanatas}. In \cite{alrabadi_PSK}, by decomposing the radiation patterns of a compact switched parasitic antenna in the far-field into a natural basis that inherently exists in the array factor itself, PSK modulation of any order could be supported. The methodology was generalized in \cite{barousis} using Gram-Schmidt decomposition for constructing a set of orthogonal functions. Unfortunately, both \cite{alrabadi_PSK} and \cite{barousis} rely on restrictive or unrealistic theoretical assumptions, such as  requiring the consideration of ideal dipole antennas not representative of modern mobile terminal antennas, while approximating their far-field by 2D horizontal cuts for the modulating-loads calculations. In addition, switching the reactive loading following the methods proposed previously results in large dynamic variation of the driving antenna input impedance, hence low matching efficiency during most symbol transmissions. This seriously undermines the interest in the novel single-radio MIMO approach since the need for a symbol-rate dynamic matching network would offset the main benefit of the single-radio MIMO system, namely, reduced RF hardware complexity.

Other key steps towards the realization of the single-radio MIMO concept were recently reported in \cite{alrabadi_tap,julien_eumc,my_tap}. In \cite{alrabadi_tap}, the derivation of the orthogonal basis functions from mirrored beam patterns of a symmetric switched parasitic antenna was suggested, based on which the first fully-operational single-radio MIMO system was designed in \cite{julien_eumc} and experimentally demonstrated in \cite{alrabadi_tap}. Thereafter, the first integrated antenna solution for implementing the single-radio MIMO concept in real small portable devices was presented in \cite{my_tap} where instead of a set of dipole and monopole radiators, a compact multi-port built-in radiating structure was used. However, while applying such an approach to BPSK signaling is straightforward, scaling to higher order PSK modulation is not possible.

In this context, this work proposes an efficient single-radio MIMO strategy which enables multiplexing \textit{higher order} PSK data streams with a realistic single-feed reconfigurable antenna. This is achieved by reconsidering the signal-space multiplexing approach and viewing it as a multi-layer analogue precoding. In the proposed approach, multiplexing of basis radiation patterns is replaced with multiplexing of basis vectors as the basis vectors can be precisely obtained without extracting complex far-field radiation patterns, so avoiding tedious far-field calculations when deriving the multiplexing relation. The approach makes the use of only variable passive loads for pattern reconfiguration, thereby reducing complexity, power consumption, and potential stability issues. Moreover, the proposed approach ensures a constant input reflection coefficient of the single-feed reconfigurable antenna independently of the two data streams, consequently obviating the use of symbol-rate dynamic matching networks. The procedure is illustrated by an antenna design example supporting the single-feed transmission of two QPSK data streams.

Notation: In the following, boldface lower-case and uppercase characters denote vectors and matrices, respectively. The operators $(\cdot)^*$, $(\cdot)^\text{T}$, and $(\cdot)^\text{H}$ designate complex conjugate, transpose, and complex conjugate transpose (Hermitian) operators, respectively. The notation $\textbf{I}_N$ indicates an identity matrix of size $N\times N$. $(\cdot)_{ij}$ returns the $\{i,j\}$ entry of the enclosed matrix and $(\cdot)_i$ returns the $i$th element of the enclosed vector whereas $|\cdot|$ returns the absolute value.  The operator $\in$ indicates that the (random) variable belongs to a certain set of numbers.

\section{Background and Theory} \label{section2}
\begin{figure}[b!]
\centering
\includegraphics[width=5.4in]{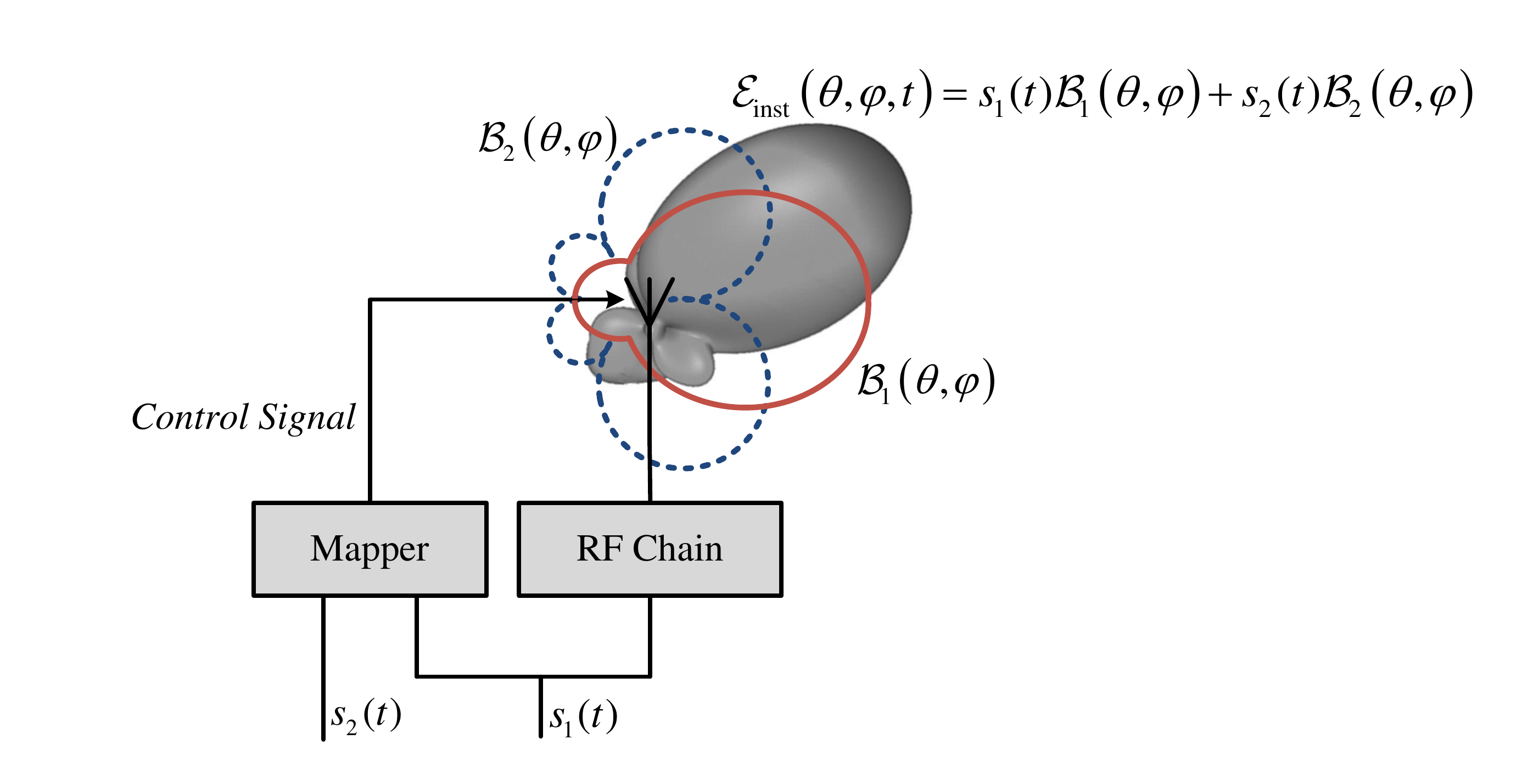}
\caption{Symbolic representation of the beam-space MIMO concept. The orthogonal basis patterns ensure decorrelation of the MIMO coefficients in a rich-scattered channel, and allows decoding the multiplexed symbols at the receiver.}
\label{fig:fig1}
\end{figure}

The main idea in the beam-space MIMO concept is to modulate some of the MIMO data sub-streams directly onto the antenna far-field \cite{kalis2008}. For this purpose, a reconfigurable antenna should be devised such that its instantaneous radiation pattern is decomposable at any instant of time as follows:
\begin{equation}\label{eq:eq1}
\CMcal{E}_\text{inst}( {\theta{,}\varphi ,t}) = {s_1}( t )\CMcal{B}_1( {\theta{,}\varphi } ) + {s_2}( t )\CMcal{B}_2( {\theta{,}\varphi } )
\end{equation}
where $\CMcal{B}_1( {\theta{,}\varphi } )$ and $\CMcal{B}_2( {\theta{,}\varphi } )$ form an orthogonal basis in the beam-space domain of the antenna, and ${s_1}( t )$ and ${s_2}( t )$ are independent complex symbols from the signal constellation diagram. In this case, as shown in Fig. \ref{fig:fig1}, ${s_1}( t )$ and ${s_2}( t )$  are driven to different virtual antennas in the beam-space domain of the reconfigurable antenna, modulating the orthogonal basis patterns. Under rich scattering conditions, a receiver equipped with multiple independent antennas attached to multiple independent RF chains may decode the transmitted mixture of signals by estimating the receive antenna responses to the corresponding basis \cite{alrabadi_tap}.

Here, we introduce a powerful multiplexing approach which makes the implementation of (\ref{eq:eq1}) possible for any PSK modulation using a compact single-feed antenna with only passive loads embedded. In the following, we start by briefly recalling theoretical background on multiport antennas that are necessary for the new calculations. Then, the mathematical developments related to the novel proposed methodology are described in detail.
\subsection{Recall on Theory of Multiport Antennas}
\vspace{-1em}
An \emph{N}-port radiator can be fully described by an \emph{N}-by-\emph{N} scattering matrix ${\boldsymbol{\CMcal{S}}}$ and \emph{N} embedded radiation patterns (also called active port patterns). Assuming a linear media, the scattering parameters are defined as
\begin{equation}\label{eq:eq2}
{\boldsymbol{b}} = {\boldsymbol{\CMcal{S}a}}
\end{equation}
where ${a_n}=(\boldsymbol{a})_n$ and ${b_n}=(\boldsymbol{b})_n$, $n \in \{ {0,{\rm{1}}, \ldots ,N{-}{\rm{1}}} \}$, are the incident and reflected power waves at port \emph{n}, respectively. Generally, the power waves can be defined so that a unit magnitude corresponds to a unit power level of the incident and reflected waves, i.e.
\begin{equation}\label{eq:eq3}
\begin{array}{l}
\CMcal{P}_{\text{inc},n} = {\left| {{a_n}} \right|^2}\\
{\CMcal{P}_{\text{ref},n}} = {\left| {{b_n}} \right|^2} .
\end{array}
\end{equation}
The embedded radiation pattern of port \emph{i}, denoted by $\CMcal{E}_i( {\theta{,}\varphi } )$, is defined as the radiation pattern obtained when driving port \emph{i} with a unit power level and terminating all other ports with a reference impedance $Z_0$ \cite{pozar}, namely, when ${a_i}= 1\,{{\text{W}}^{1/2}}$ and ${a_{n \ne i}} = 0$. Notice that the energy absorbed inside the system or carried away from the system in reflected waves and absorbed in terminating matched loads has been already considered in the definition of the embedded pattern. In the following, we recall the analytical expressions of the beam-coupling coefficients utilized later for defining a set of orthogonal basis vectors.

The total power incident onto the radiator is equal to the sum of the powers incident at the individual ports, thus using (\ref{eq:eq3})
\begin{equation}\label{eq:eq4}
{\CMcal{P}_{\text{inc},\text{tot}}} = \sum\limits_{n = 0}^{N - 1} {\CMcal{P}_{\text{inc},n}}  = \sum\limits_{n = 0}^{N - 1} {{{\left| {{a_n}} \right|}^2}}  = {{\boldsymbol{a}}^\text{H}}{\boldsymbol{a}} .
\end{equation}
Similarly, the total power reflected back from the radiator is the sum of the powers reflected back at the individual ports, thus using (\ref{eq:eq2}) and (\ref{eq:eq3})
\begin{eqnarray} \label{eq:eq5}
{\CMcal{P}_{\text{ref},\text{tot}}}  &=& \sum\limits_{n = 0}^{N - 1} {{\CMcal{P}_{\text{ref},n}}}  = \sum\limits_{n = 0}^{N - 1} {{{\left| {{b_n}} \right|}^2}} \nonumber  \\
   &=& {{\boldsymbol{b}}^\text{H}}{\boldsymbol{b}} = {{\boldsymbol{a}}^\text{H}}{{\boldsymbol{\CMcal{S}}}^\text{H}}{\boldsymbol{\CMcal{S}a}} .
\end{eqnarray}
Further, the total radiated field can also be expressed by superposition as a linear combination of the embedded patterns \cite{stein}, i.e.
\begin{equation}\label{eq:eq6}
\CMcal{E}_{\text{tot}}( {\theta{,}\varphi } ) = \sum\limits_{n = 0}^{N - 1} {{a_n}\CMcal{E}_n( {\theta{,}\varphi })}
\end{equation}
where the units of $\CMcal{E}_{\text{tot}}( {\theta{,}\varphi } )$ and $\CMcal{E}_{n}( {\theta{,}\varphi } )$ are V/m and V/m/$\text{W}^{1/2}$, respectively.
From (\ref{eq:eq6}), the total power radiated from the radiator is
\begin{align}\label{eq:eq7}
  {\CMcal{P}_{\text{rad},\text{tot}}} &= \frac{1}{2}{\eta _0}\int\!\!\!\!\!\int {\CMcal{E}_{\text{tot}}( {\theta{,}\varphi } ) \cdot {\CMcal{E}_{\text{tot}}}^*( {\theta{,}\varphi } )\,dS} \nonumber \\
   &= \sum\limits_{m = 1}^{N - 1} {\sum\limits_{n = 1}^{N - 1} a_n^*\left[ \frac{1}{2}{\eta _0}\int\!\!\!\!\!\int {\CMcal{E}_{m}( {\theta{,}\varphi } ) \cdot {\CMcal{E}_{n}}^*( {\theta{,}\varphi } )\,dS}\right]a_m}\nonumber \\
   &= \sum\limits_{m = 1}^{N - 1} {\sum\limits_{n = 1}^{N - 1} {{a_n}^ * {\chi _{nm}}\,} } {a_m} = {{\boldsymbol{a}}^\text{H}}{\boldsymbol{\CMcal{X} a}}
\end{align}
where
\begin{equation}\label{eq:eq8}
{\chi _{nm}}\, = \frac{1}{2}{\eta _0}\int\!\!\!\!\!\int {{\CMcal{E}_m}( {\theta{,}\varphi } ) \cdot {\CMcal{E}_n}^*( {\theta{,}\varphi } )dS}
\end{equation}
is defined as the beam-coupling coefficient between \emph{n}th and \emph{m}th embedded patterns \cite{stein}, ${\eta _0}$ is the free-space characteristic admittance, and  $dS = r^2sin\theta d\theta d\varphi$. It can be seen from (\ref{eq:eq8}) that  ${\chi _{nm}} = {\chi _{mn}}^ *$.

According to (\ref{eq:eq8}), the set of beam-coupling coefficients are typically obtained through tedious far-field calculations. This may increase the computational complexity associated with the optimization procedure of single-radio MIMO systems. However, when the thermal losses in the radiator materials are negligible, the knowledge of the scattering parameters suffices for calculating the beam-coupling coefficients. Indeed, assuming negligible loss in the antenna materials, energy conservation implies that the total radiated power is equal to the difference between the total incident and reflected powers. Using (\ref{eq:eq4}), (\ref{eq:eq5}) and (\ref{eq:eq7}), this leads to
\begin{equation}\label{eq:eq9}
{{\boldsymbol{a}}^\text{H}}{\boldsymbol{\CMcal{X} a}} = {{\boldsymbol{a}}^\text{H}}{\boldsymbol{a}} - {{\boldsymbol{a}}^\text{H}}{{\boldsymbol{\CMcal{S}}}^\text{H}}{\boldsymbol{\CMcal{S}a}}.
\end{equation}
Since (\ref{eq:eq9}) is valid for all complex $\boldsymbol{a}$, it can be simply demonstrated that (see Appendix \ref{appendix:II})
\begin{equation}\label{eq:eq10}
{\boldsymbol{\CMcal{X}}} = {\textbf{I}} - {{\boldsymbol{\CMcal{S}}}^\text{H}}{\boldsymbol{\CMcal{S}}}.
\end{equation}
Therefore, the beam-coupling coefficients between the embedded patterns of a lossless radiator can be expressed in terms of the scattering parameters only,
\begin{equation}\label{eq:eq11}
{\chi _{nm}}\, =  - \sum\limits_{p = 0}^{N - 1} {{\CMcal{S}_{pn}}^*{\CMcal{S}_{pm}}} ,\quad n \ne m
\end{equation}
thus eliminating the need for cumbersome far-field calculations. Similarly, the total power radiated in the far-field region caused by a unit power incident on port \emph{n} can also be derived as a function of the scattering parameters,
\begin{equation}\label{eq:eq12}
{\CMcal{P}_{\text{rad},n}} = {\chi _{nn}} = 1 - \sum\limits_{p = 0}^{N - 1} {{{\left| {{\CMcal{S}_{pn}}} \right|}^2}}.
\end{equation}
\subsection{Reconfigurable Antennas with Mirrored Beam Patterns}
\begin{figure}[t!]
\centering
\subfigure[]{\includegraphics[width=3.3in]{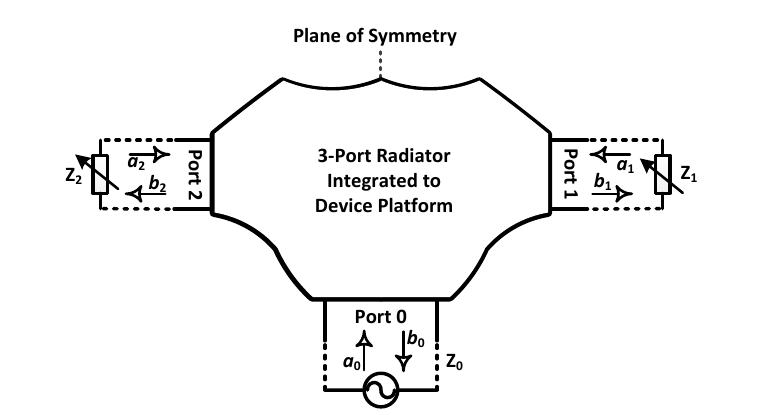} \label{fig:fig2a}}
\subfigure[]{\includegraphics[width=2.7in]{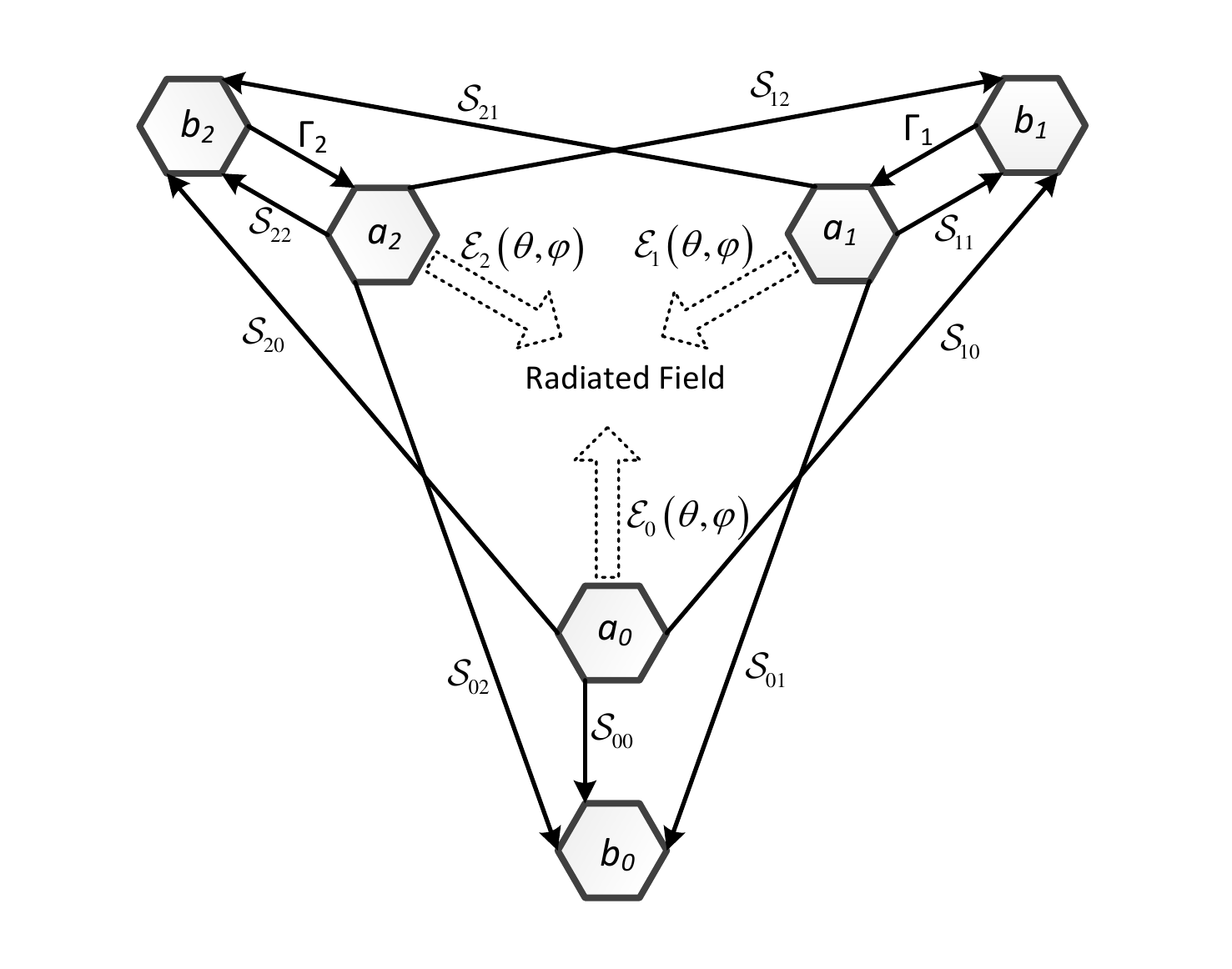} \label{fig:fig2b}}
\caption{(a) A symmetric three-port antenna system. The central port is the active input, whereas the other two are control ports, loaded by passive loads. The source and reference impedances are assumed to be identical. Note that the single-port structure (i.e. when two control ports are terminated) is not generally symmetrical. (b) Equivalent signal flow of the radiator. The incident and reflected power waves are denoted by $a_i$ and $b_i$, respectively. Obviously by symmetry, $\CMcal{S}_{11} = \CMcal{S}_{22}$ and $\CMcal{S}_{01}=\CMcal{S}_{02}$ . }
\label{fig:fig2}
\vspace{-1em}
\end{figure}

Fig. \ref{fig:fig2a} shows a single-feed reconfigurable antenna system comprising a symmetric three-port radiator and two variable loads $Z_1$ and $Z_2$ connected to the radiator passive ports (also referred to as control ports). The control ports can be either mounted on the main radiating structure or on separate radiators parasitically coupled to the main radiator \cite{alrabadi_tap,julien_eumc,my_tap}. In this section, we first briefly recall the analytic expressions of the antenna reflection coefficient and the antenna radiated field in terms of the radiator parameters and the variable loads. This will allow us later to define our set of orthogonal basis vectors for single-radio multiplexing.

Based on the approach in \cite{petit}, the antenna system of Fig. \ref{fig:fig2a} can be modeled by the signal flow graph of Fig. \ref{fig:fig2b} where
\begin{equation}\label{eq:eq13}
{\Gamma _k} = {{\left( {{Z_k} - {Z_0}} \right)} \mathord{\left/
 {\vphantom {{\left( {{Z_k} - {Z_0}} \right)} {\left( {{Z_k} + {Z_0}} \right)}}} \right.
 \kern-\nulldelimiterspace} {\left( {{Z_k} + {Z_0}} \right)}}\quad k \in \left\{ {1,2} \right\}
\end{equation}
is the reflection coefficient at port \emph{k}. For the sake of simplicity, the source impedance $Z_0$ is chosen to be equal to the reference impedance of the scattering parameters. Using Mason’s rule \cite{mason}, the total reflection coefficient at the central active RF input of this symmetrical structure is derived as
\begin{equation}\label{eq:eq14}
{\Gamma _{\text{tot}}} = {{b{}_0} \mathord{\left/
 {\vphantom {{b{}_0} {{a_0}}}} \right.
 \kern-\nulldelimiterspace} {{a_0}}} = {\CMcal{S}_{00}} + {\CMcal{S}_{01}}\left( {\ell _1^{} + \ell _2^{}} \right)
\end{equation}
where
\begin{subequations}\label{eq:eq15}
\begin{align}
\ell _1^{} &= {\Gamma _1}\frac{{{b_1}}}{{{a_0}}} \nonumber\\
&= {\Gamma _1}{\CMcal{S}_{01}}\frac{{1 - {\Gamma _2}\left( {{\CMcal{S}_{11}} - {\CMcal{S}_{21}}} \right)}}{{1 - {\CMcal{S}_{11}}\left( {{\Gamma _1} + {\Gamma _2}} \right) + {\Gamma _1}{\Gamma _2}\left( {\CMcal{S}_{11}^2 - \CMcal{S}_{21}^2} \right)}} \\
\ell _2^{} &= {\Gamma _2}\frac{{{b_2}}}{{{a_0}}} \nonumber\\
&= {\Gamma _2}{\CMcal{S}_{01}}\frac{{1 - {\Gamma _1}\left( {{\CMcal{S}_{11}} - {\CMcal{S}_{21}}} \right)}}{{1 - {\CMcal{S}_{11}}\left( {{\Gamma _1} + {\Gamma _2}} \right) + {\Gamma _1}{\Gamma _2}\left( {\CMcal{S}_{11}^2 - \CMcal{S}_{21}^2} \right)}}  .
\end{align}
\end{subequations}
Similarly, the antenna total radiated field when exciting the active port with a unit power can be expressed as a linear combination of the three embedded patterns:
\begin{equation}\label{eq:eq16}
{\CMcal{E}_\text{unit}^{\text{}}}( {\theta{,}\varphi } ) = \boldsymbol{v}_\text{inst}\boldsymbol{\CMcal{E}}_\text{emb}^\text{T}
\end{equation}
where
\begin{equation}\label{eq:eq17}
\boldsymbol{\CMcal{E}}_\text{emb} = \left[ {\begin{array}{*{20}{c}}
{{\CMcal{E}_0}( {\theta{,}\varphi } )}&{{\CMcal{E}_1}( {\theta{,}\varphi } )}&{{\CMcal{E}_2}( {\theta{,}\varphi } )}
\end{array}} \right]
\end{equation}
is the vector of the radiator embedded patterns, and
\begin{equation}\label{eq:eq18}
\boldsymbol{v}_\text{inst}^{} = \left[ {\begin{array}{*{20}{c}}
1&{\ell _1^{}}&{\ell _2^{}}
\end{array}} \right]
\end{equation}
is defined here as the unit instantaneous pattern vector. Let us emphasize here that all instantaneous radiation patterns as well as the three embedded patterns are \emph{complex vectorial }angular functions.

Due to the symmetry of the radiator, the permutation of the loads at the control ports will mirror the antenna radiation pattern with respect to the plane of symmetry, while the total power radiated in the far-field will remain constant. This feature is employed in the next section, where the desired basis for single-radio MIMO transmission is defined.

\subsection{Definition of Orthogonal Basis Vectors}
 As discussed at the beginning of Section \ref{section2}, the beam-space MIMO concept requires the decomposition of the antenna instantaneous radiation pattern into proper weighted sum of orthogonal basis patterns. Here, we demonstrate that for the reconfigurable antenna system symbolically represented in Fig. \ref{fig:fig2a}, the definition of angular functions $\CMcal{B}_1$ and $\CMcal{B}_2$ as
\begin{subequations}\label{eq:eq19}
\begin{align}
{\CMcal{B}_1}( {\theta{,}\varphi } ) &= \frac{{\CMcal{E}_\text{unit}^{\{ {{Z_\text{II}},{Z_\text{I}}} \}}( {\theta{,}\varphi } ) + \CMcal{E}_\text{unit}^{\{ {{Z_\text{I}},{Z_\text{II}}} \}}( {\theta{,}\varphi } )}}{2}\\
{\CMcal{B}_2}( {\theta{,}\varphi } ) &= \frac{{\CMcal{E}_\text{unit}^{\{ {{Z_\text{II}},{Z_\text{I}}} \}}( {\theta{,}\varphi } ) - \CMcal{E}_\text{unit}^{\{ {{Z_\text{I}},{Z_\text{II}}} \}}( {\theta{,}\varphi } )}}{2}
\end{align}
\end{subequations}
creates the desired orthogonal basis for decomposing the instantaneous radiation fields. In (\ref{eq:eq19}), the superscript notation determines two distinct system states, namely

\begin{itemize}
  \item State $\{ {Z_\text{I},Z_\text{II}} \}$ where the loads $Z_\text{I}$ and $Z_\text{II}$ are connected to the ports 1 and 2, respectively, i.e. $\Gamma _1^{\{ {{Z_\text{I}},{Z_\text{II}}} \}} = {\Gamma _\text{I}}$ and $\Gamma _2^{\{ {{Z_\text{I}},{Z_\text{II}}} \}} = {\Gamma _\text{II}}$;
  \item State $\{ {Z_\text{II},Z_\text{I}} \}$ where the loads $Z_\text{I}$ and $Z_\text{II}$ are connected to the ports 2 and 1, respectively, i.e. $\Gamma _1^{\{ {{Z_\text{II}},{Z_\text{I}}} \}} = {\Gamma _\text{II}}$ and $\Gamma _2^{\{ {{Z_\text{II}},{Z_\text{I}}} \}} = {\Gamma _\text{I}}$.
\end{itemize}
In the following, we start by writing the analytical expressions of the basis patterns in terms of the radiator embedded radiation patterns. Then, we show that $\CMcal{B}_1$ and $\CMcal{B}_2$ are orthogonal.
\\
\indent
When the system state is changed, it is obvious from (\ref{eq:eq13}) that the reflection coefficients at the control ports are permuted: $\Gamma _1^{\{ {{Z_\text{I}},{Z_\text{II}}} \}} = \Gamma _2^{\{ {{Z_\text{II}},{Z_\text{I}}} \}} = {\Gamma _\text{I}}$  and $\Gamma _2^{\{ {{Z_\text{I}},{Z_\text{II}}} \}} = \Gamma _1^{\{ {{Z_\text{II}},{Z_\text{I}}} \}} = {\Gamma _\text{II}}$ . Accordingly, the coefficients defined by (\ref{eq:eq15}) are updated as follows:
\begin{subequations}\label{eq:eq20}
\begin{align}
\ell _1^{\{ {{Z_\text{I}},{Z_\text{II}}} \}} &= \ell _2^{\{ {{Z_\text{II}},{Z_\text{I}}}\}} \nonumber\\
&= \frac{{\Gamma_\text{I}}{\CMcal{S}_{01}}[1 - {\Gamma_\text{II}}( {{\CMcal{S}_{11}} - {\CMcal{S}_{21}}})]}{{1 - {\CMcal{S}_{11}}( {{\Gamma_\text{I}} + {\Gamma_\text{II}}}) + {\Gamma_\text{I}}{\Gamma_\text{II}}( {\CMcal{S}_{11}^2 - \CMcal{S}_{21}^2})}}\\[2ex]
\ell _2^{\{ {{Z_\text{I}},{Z_\text{II}}}\}} &= \ell _1^{\{ {{Z_\text{II}},{Z_\text{I}}}\}} \nonumber\\
&= \frac{{\Gamma_\text{II}}{\CMcal{S}_{01}}[1 - {\Gamma_\text{I}}({{\CMcal{S}_{11}} - {\CMcal{S}_{21}}})]}{{1 - {\CMcal{S}_{11}}( {{\Gamma_\text{I}} + {\Gamma_\text{II}}} ) + {\Gamma_\text{I}}{\Gamma_\text{II}}( {\CMcal{S}_{11}^2 - \CMcal{S}_{21}^2} )}}
\end{align}
\end{subequations}
Using (\ref{eq:eq16}) and (\ref{eq:eq20}), the antenna total radiated fields for two given system states are derived as
\begin{subequations}\label{eq:eq21}
\begin{align}
\CMcal{E}_\text{unit}^{\{ {{Z_\text{I}},{Z_\text{II}}} \}}( {\theta{,}\varphi } ) = \boldsymbol{v}_\text{inst}^{\{ {{Z_\text{I}},{Z_\text{II}}} \}}\boldsymbol{\CMcal{E}}_\text{emb}^\text{T} \\[2ex]
\CMcal{E}_\text{unit}^{\{ {{Z_\text{II}},{Z_\text{I}}} \}}( {\theta{,}\varphi } ) = \boldsymbol{v}_\text{inst}^{\{ {{Z_\text{II}},{Z_\text{I}}} \}}\boldsymbol{\CMcal{E}}_\text{emb}^\text{T}
\end{align}
\end{subequations}
where
\begin{subequations}\label{eq:eq22}
\begin{align}
\boldsymbol{v}_\text{inst}^{\{ {{Z_\text{I}},{Z_\text{II}}} \}} &= \left[ {\begin{array}{*{20}{c}}
1&{\ell _1^{\{ {{Z_\text{I}},{Z_\text{II}}} \}}}&{\ell _2^{\{ {{Z_\text{I}},{Z_\text{II}}} \}}}
\end{array}} \right]\\[2ex]
\boldsymbol{v}_\text{inst}^{\{ {{Z_\text{II}},{Z_\text{I}}} \}} &= \left[ {\begin{array}{*{20}{c}}
1&{\ell _2^{\{ {{Z_\text{I}},{Z_\text{II}}} \}}}&{\ell _1^{\{ {{Z_\text{I}},{Z_\text{II}}} \}}}
\end{array}} \right]
\end{align}
\end{subequations}
are corresponding unit instantaneous pattern vectors. Since $\CMcal{E}_2$ is a mirrored version of $\CMcal{E}_1$ with regard to the plane of symmetry, and $\CMcal{E}_0$ is symmetric with respect to the same plane, it can be inferred from (\ref{eq:eq21}) that $\CMcal{E}_\text{unit}^{\{ {{Z_\text{II}},{Z_\text{I}}} \}}$ is a mirrored version of $\CMcal{E}_\text{unit}^{\{ {{Z_\text{I}},{Z_\text{II}}} \}}$ regarding the plane of symmetry.
Now, using (\ref{eq:eq19}), (\ref{eq:eq21}) and (\ref{eq:eq22}), the basis functions are directly rewritten in terms of the three embedded patterns as
\begin{subequations}\label{eq:eq23}
\begin{align}
{\CMcal{B}_1}( {\theta ,\varphi } ) &= \boldsymbol{v}_{{\CMcal{B}_1}}^{\{ {{Z_\text{I}},{Z_\text{II}}}\}}\boldsymbol{\CMcal{E}}_\text{emb}^\text{T}\\[2ex]
{\CMcal{B}_2}( {\theta ,\varphi } ) &= \boldsymbol{v}_{{\CMcal{B}_2}}^{\{ {{Z_\text{I}},{Z_\text{II}}}\}}\boldsymbol{\CMcal{E}}_\text{emb}^\text{T}
\end{align}
\end{subequations}
where
\begin{subequations}\label{eq:eq24}
\begin{align}
\boldsymbol{v}_{{\CMcal{B}_1}}^{\{ {{Z_\text{I}},{Z_\text{II}}}\}} &= \left[ {\begin{array}{*{20}{c}}
1&{\ell _{\CMcal{B}_1}^{\{ {{Z_\text{I}},{Z_\text{II}}} \}}}&{\ell _{\CMcal{B}_1}^{\{ {{Z_\text{I}},{Z_\text{II}}} \}}}
\end{array}} \right]\\[2ex]
\boldsymbol{v}_{{\CMcal{B}_2}}^{\{ {{Z_\text{I}},{Z_\text{II}}}\}} &= \left[ {\begin{array}{*{20}{c}}
0&{\ell _{\CMcal{B}_2}^{\{ {{Z_\text{I}},{Z_\text{II}}} \}}}&{ - \ell _{\CMcal{B}_2}^{\{ {{Z_\text{I}},{Z_\text{II}}} \}}}
\end{array}} \right]
\end{align}
\end{subequations}
are the basis vectors and
\begin{subequations}\label{eq:eq25}
\begin{align}
\ell _{\CMcal{B}_1}^{\{ {{Z_\text{I}},{Z_\text{II}}} \}} &= \frac{{\ell _1^{\{ {{Z_\text{I}},{Z_\text{II}}} \}} + \ell _2^{\{ {{Z_\text{I}},{Z_\text{II}}} \}}}}{2} \nonumber \\
&= \frac{\frac{1}{2}\CMcal{S}_{01}\left[{{\Gamma_\text{I}} + {\Gamma_\text{II}} - 2{\Gamma_\text{I}}{\Gamma_\text{II}}\left( {{\CMcal{S}_{11}} - {\CMcal{S}_{21}}} \right)}\right]}{{1 - {\CMcal{S}_{11}}\left( {{\Gamma_\text{I}} + {\Gamma_\text{II}}} \right) + {\Gamma_\text{I}}{\Gamma_\text{II}}\left( {\CMcal{S}_{11}^2\!-\!\CMcal{S}_{21}^2} \right)}}\\[2ex]
\ell _{\CMcal{B}_2}^{\{ {{Z_\text{I}},{Z_\text{II}}} \}} &= \frac{{\ell _2^{\{ {{Z_\text{I}},{Z_\text{II}}} \}} - \ell _1^{\{ {{Z_\text{I}},{Z_\text{II}}} \}}}}{2}\nonumber\\
&= \frac{\frac{1}{2}\CMcal{S}_{01}\left[{{\Gamma_\text{II}} - {\Gamma_\text{I}}}\right]}{{1 - {\CMcal{S}_{11}}\left( {{\Gamma_\text{I}} + {\Gamma_\text{II}}} \right) + {\Gamma_\text{I}}{\Gamma_\text{II}}\left( {\CMcal{S}_{11}^2\!-\!\CMcal{S}_{21}^2} \right)}} \cdot
\end{align}
\end{subequations}
Now, the beam-coupling coefficient between $\CMcal{B}_1$ and $\CMcal{B}_2$ can be calculated using (\ref{eq:eq8}) as
\begin{align}\label{eq:eq26}
{\chi_{{\CMcal{B}_1}\!{\CMcal{B}_2}}}&= \frac{{{\eta _0}}}{2}\int\!\!\!\!\!\int {{\CMcal{B}_2}( {\theta{,}\varphi }) \cdot \CMcal{B}_1^*( {\theta{,}\varphi } )dS} \nonumber\\
&= \frac{{{\eta _0}}}{2}\int\!\!\!\!\!\int {\left[ {\boldsymbol{v}_{{\CMcal{B}_2}}^{\{ {{Z_\text{I}},{Z_\text{II}}} \}}\boldsymbol{\CMcal{E}}_\text{emb}^\text{T}} \right] \cdot {{\left[ {\boldsymbol{v}_{{\CMcal{B}_1}}^{\{ {{Z_\text{I}},{Z_\text{II}}} \}}\boldsymbol{\CMcal{E}}_\text{emb}^\text{T}} \right]}^*}dS} \nonumber\\
&= \frac{{{\eta _0}}}{2}\int\!\!\!\!\!\int {\ell _{{\CMcal{B}_2}}^{\{ {{Z_\text{I}},{Z_\text{II}}} \}}\left[ {{\CMcal{E}_1} - {\CMcal{E}_2}} \right]}\nonumber\\
 &\qquad\qquad\cdot{\left[ {{\CMcal{E}_0} + \ell _{{\CMcal{B}_1}}^{\{ {{Z_\text{I}},{Z_\text{II}}} \}}\left( {{\CMcal{E}_1} + {\CMcal{E}_2}} \right)} \right]^*}dS \nonumber\\
&= \frac{{{\eta _0}}}{2}\ell _{{\CMcal{B}_2}}^{\{ {{Z_\text{I}},{Z_\text{II}}} \}}\left( {\chi_{01}}-{\chi _{02}}\right) \nonumber\\
&\quad+\!\frac{{{\eta _0}}}{2}\ell _{{\CMcal{B}_2}}^{\{\!{Z_\text{I}}{,}{Z_\text{II}\!}\}} \ell {{_{{\CMcal{B}_1}}^{\{\!{Z_\text{I}}{,}{Z_\text{II}\!}\}}}^{ * }}\!( {{\chi _{11}}\! -\! {\chi _{22}}\! +\! {\chi _{21}}\! -\! {\chi _{12}}} ) .
\end{align}
Since by symmetry of the radiator, ${\chi _{0{\rm{1}}}} = {\chi _{0{\rm{2}}}}$, ${\chi _{11}} = {\chi _{22}}$ and ${\chi _{21}} = {\chi _{12}}$,
\begin{equation}\label{eq:eq27}
{\chi _{{\CMcal{B}_1}\!{\CMcal{B}_2}}} = 0
\end{equation}
which concludes the demonstration by showing that $\CMcal{B}_1$ and $\CMcal{B}_2$ as defined in (\ref{eq:eq19}) form an orthogonal basis. Moreover, we can show that the basis vectors are also orthogonal as their dot product is zero:
\begin{align}\label{eq:eq28}
\boldsymbol{v}_{{\CMcal{B}_1}}^{\{ {{Z_\text{I}},{Z_\text{II}}} \}} \cdot \boldsymbol{v}_{{\CMcal{B}_2}}^{\{ {{Z_\text{I}},{Z_\text{II}}} \}} = 0 .
\end{align}

It is worth noting that the orthogonality of $\CMcal{B}_1$ and $\CMcal{B}_2$ is valid regardless of the impedances $Z_\text{I}$ and $Z_\text{II}$ (hereafter also called the basis impedances). However, their values affect the total reflection coefficient at the single active port as well as the powers $\CMcal{P}_{\CMcal{B}_1}$ and $\CMcal{P}_{\CMcal{B}_2}$ radiated in the far-field by the basis patterns. Since for open-loop MIMO operation, a balanced power distribution between the multiple streams is ideally desired, here we define
\begin{equation}\label{eq:eq29}
r = {{{\CMcal{P}_{{\CMcal{B}_1}}}} \mathord{\left/
 {\vphantom {{{\CMcal{P}_{{\CMcal{B}_1}}}} {{\CMcal{P}_{{\CMcal{B}_2}}}}}} \right.
 \kern-\nulldelimiterspace} {{\CMcal{P}_{{\CMcal{B}_2}}}}}
\end{equation}
as the power imbalance ratio between the basis patterns. We will show later in Section \ref{section3} the importance of this factor in the design of single-radio MIMO systems. In general, the power imbalance ratio is a function of all antenna input parameters, including the radiator embedded patterns. However, as shown in Appendix \ref{appendix:I}, the power imbalance can also be expressed in terms of the scattering parameters and the basis impedances $Z_\text{I}$ and $Z_\text{II}$ as long as the ohmic and dielectric losses in the radiator materials are negligible.

\subsection{Beam-Space Multiplexing Technique}
\vspace{-1em}

In the previous section we proved the existence of a natural orthogonal basis for the single-feed pattern-reconfigurable antenna system shown in Fig. \ref{fig:fig2a}. We follow here by demonstrating an efficient approach that makes such an antenna system capable of multiplexing two data streams of any modulation order.

The proposed technique consists in the proper selection of the set of the impedances $Z_1$ and $Z_2$ at the control ports such that the antenna instantaneous radiated field satisfies (\ref{eq:eq1}) for any combination of two data streams $s_1(t)$ and $s_2(t)$. In other words, for mapping each arbitrary symbol combination of $\{s_1{,}s_2\}$ from the considered signal constellation diagram on the basis functions already defined in (\ref{eq:eq23}) and enabling single-radio spatial multiplexing, we need to find the loading values $Z_1^{\left\{ {{s_1},{s_2}} \right\}}$ and $Z_2^{\left\{ {{s_1},{s_2}} \right\}}$ such that,
\begin{align}\label{eq:eq30}
\CMcal{E}_\text{inst}^{}( {\theta{,}\varphi{,}{s_1}{,}{s_2}} ) &= {s_1}{\CMcal{B}_1}( {\theta{,}\varphi } ) + {s_2}{\CMcal{B}_2}( {\theta{,}\varphi }) \nonumber \\
&= \left[{s_1}\boldsymbol{v}_{{\CMcal{B}_1}}^{\{ {{Z_\text{I}},{Z_\text{II}}} \}} + {s_2}\boldsymbol{v}_{{\CMcal{B}_2}}^{\{ {{Z_\text{I}},{Z_\text{II}}} \}}\right]\boldsymbol{\CMcal{E}}_\text{emb}^\text{T} .
\end{align}
where $\CMcal{E}_\text{inst}( {\theta{,}\varphi{,}{s_1}{,}{s_2}} )$ is the antenna instantaneous radiated field for the symbol pair $\{s_1{,}s_2\}$. On the other hand, $\CMcal{E}_\text{inst}( {\theta{,}\varphi{,}{s_1}{,}{s_2}} )$ can generally be written as the multiplication of the antenna total radiated field for a unit power excitation $\CMcal{E}_\text{unit}( {\theta{,}\varphi{,}{s_1}{,}{s_2}} )$ and the signal applied to the antenna system at its single active input, which we define as $s_\text{in}( {{s_1}{,}{s_2}} )$, thus
\begin{align}\label{eq:eq31}
{\CMcal{E}_\text{inst}}( {\theta{,}\varphi{,}{s_1}{,}{s_2}} )\!&=\!{s_\text{in}}( {{s_1},{s_2}} )\CMcal{E}_\text{unit}^{\{ {Z_1^{\{ {{s_1},{s_2}} \}},Z_2^{\{ {{s_1},{s_2}} \}}} \}}( {\theta{,}\varphi } ) \nonumber\\
&={s_\text{in}}( {{s_1},{s_2}} )\boldsymbol{v}_\text{inst}^{\{ {Z_1^{\{ {{s_1},{s_2}} \}},Z_2^{\{ {{s_1},{s_2}} \}}} \}}\boldsymbol{\CMcal{E}}_\text{emb}^\text{T}
\end{align}
where according to (\ref{eq:eq18})
\begin{equation}\label{eq:eq32}
\boldsymbol{v}_\text{inst}^{\{\!{Z_1^{\{\! {{s_1},{s_2}}\! \}}\!,Z_2^{\{\! {{s_1},{s_2}}\! \}}} \!\}}\!\!\!=\!\!\! \left[\!
1\;{\ell _1^{\{\! {Z_1^{\{\! {{s_1},{s_2}} \!\}}\!,Z_2^{\{\! {{s_1},{s_2}}\! \}}}\! \}}}\;{\ell _2^{\{ \!{Z_1^{\{\! {{s_1},{s_2}}\! \}}\!,Z_2^{\{\! {{s_1},{s_2}}\! \}}} \!\}}}
 \!\right]\!\!.
\end{equation}
Combining (\ref{eq:eq30}) and (\ref{eq:eq31}), the far-field terms disappear from the equation that allows finding the unknown loads $Z_1^{\left\{ {{s_1},{s_2}} \right\}}$ and $Z_2^{\left\{ {{s_1},{s_2}} \right\}}$, thereby dispensing with cumbersome calculation of the far-field radiation patterns and replacing multiplexing of basis radiation patterns with multiplexing of basis vectors:
\begin{equation}\label{eq:eq33}
{s_\text{in}}( {{s_1},{s_2}} )\boldsymbol{v}_\text{inst}^{\{\!{Z_1^{\{\! {{s_1},{s_2}}\! \}}\!,Z_2^{\{\! {{s_1},{s_2}}\! \}}} \!\}} = {s_1}\boldsymbol{v}_{{\CMcal{B}_1}}^{\{ {{Z_\text{I}},{Z_\text{II}}} \}}\! +\! {s_2}\boldsymbol{v}_{{\CMcal{B}_2}}^{\{ {{Z_\text{I}},{Z_\text{II}}} \}}
\end{equation}
or using (\ref{eq:eq24}) and (\ref{eq:eq32}),
\begin{equation}\label{eq:eq34}
{s_\text{in}}( {{s_1},{s_2}} )\!\!\!\left[\!\!\!\!{\begin{array}{c}
1\\
{\ell _1^{\{\!{Z_1^{\{\! {{s_1},{s_2}}\! \}}\!,Z_2^{\{\! {{s_1},{s_2}}\! \}}} \!\}}}\\
{\ell _2^{\{\!{Z_1^{\{\! {{s_1},{s_2}}\! \}}\!,Z_2^{\{\! {{s_1},{s_2}}\! \}}} \!\}}}
\end{array}}\!\!\!\! \right]\!\!\! =\! {s_1}\!\!\!\left[\!\!\!\! {\begin{array}{c}
1\\
{\ell _{{\CMcal{B}_1}}^{\{ {{Z_\text{I}},{Z_\text{II}}} \}}}\\
{\ell _{{\CMcal{B}_1}}^{\{ {{Z_\text{I}},{Z_\text{II}}} \}}}
\end{array}}\!\!\!\! \right]\!\!\! +\! {s_2}\!\!\!\left[ \!\!\!\!{\begin{array}{c}
0\\
{\ell _{{\CMcal{B}_2}}^{\{ {{Z_\text{I}},{Z_\text{II}}} \}}}\\
{ - \ell _{{\CMcal{B}_2}}^{\{ {{Z_\text{I}},{Z_\text{II}}} \}}}
\end{array}}\!\!\!\! \right]\!\! .
\end{equation}
It is easily seen that a necessary condition for satisfying (\ref{eq:eq34}) is $s_\text{in}(s_1{,}s_2) = s_1$. This reveals an important practical aspect of the proposed approach: the single active port of the antenna system must be excited with one of the two data streams. In this case, (\ref{eq:eq34}) reduces to a system of two equations, allowing finding unique solutions for the unknowns $Z_1^{\left\{ {{s_1},{s_2}} \right\}}$ and $Z_2^{\left\{ {{s_1},{s_2}} \right\}}$ as functions of the radiator scattering parameters, the basis impedances $Z_\text{I}$ and $Z_\text{II}$, and the symbols pair $s_1$ and $s_2$,
\begin{equation}\label{eq:eq35}
\left[\!\!\! {\begin{array}{c}
{\ell _1^{\{\!{Z_1^{\{\! {{s_1},{s_2}}\! \}}\!,Z_2^{\{\! {{s_1},{s_2}}\! \}}} \!\}}}\\
{\ell _2^{\{\!{Z_1^{\{\! {{s_1},{s_2}}\! \}}\!,Z_2^{\{\! {{s_1},{s_2}}\! \}}} \!\}}}
\end{array}}\!\!\! \right] \!\!= \ell _{{\CMcal{B}_1}}^{\{ {{Z_\text{I}},{Z_\text{II}}} \}}\!\!\left[\!\!\! {\begin{array}{c}
1\\
1
\end{array}}\!\!\! \right] + \frac{{{s_2}}}{{{s_1}}}\ell _{{\CMcal{B}_2}}^{\{ {{Z_\text{I}},{Z_\text{II}}} \}}\!\!\left[\!\!\! {\begin{array}{c}
1\\
{ - 1}
\end{array}}\!\!\! \right]\!\! .
\end{equation}
However, as seen in (\ref{eq:eq35}), the multiplexing relation only depends on the ratio of $s_2$ and $s_1$ and not on their individual values. In other words, the same load pair is required for transmitting any symbol pair $\{s_1{,}s_2\}$ having the same ratio:
\begin{equation}\label{eq:eq36}
{s_r} = \frac{{{s_2}}}{{{s_1}}} .
\end{equation}
As a result, we simplify the notation in (\ref{eq:eq35}), replacing the superscript $\{s_1{,}s_2\}$ with $\{s_r\}$:
\begin{equation}\label{eq:eq37}
\left[\!\!\! {\begin{array}{*{20}{c}}
{\ell _1^{\{ {Z_1^{\{ {{s_r}} \}},Z_2^{\{ {{s_r}} \}}} \}}}\\
{\ell _2^{\{ {Z_1^{\{ {{s_r}} \}},Z_2^{\{ {{s_r}} \}}} \}}}
\end{array}}\!\!\! \right] = \ell _{{\CMcal{B}_1}}^{\{ {{Z_\text{I}},{Z_\text{II}}} \}}\left[ \!\!\!{\begin{array}{*{20}{c}}
1\\
1
\end{array}}\!\!\! \right] + {s_r}\ell _{{\CMcal{B}_2}}^{\{ {{Z_\text{I}},{Z_\text{II}}} \}}\left[\!\!\! {\begin{array}{*{20}{c}}
1\\
{ - 1}
\end{array}}\!\!\! \right]
\end{equation}
where $Z_1^{\{ {{s_r}} \}}$ and $Z_2^{\{ {{s_r}} \}}$ are the unknowns and using (\ref{eq:eq15})
\begin{subequations}\label{eq:eq38}
\begin{align}
{\ell _1^{\{\! {Z_1^{\{\! {{s_r}}\! \}},Z_2^{\{\! {{s_r}}\! \}}}\! \}}\!\! =\! \frac{\Gamma _1^{\{\! {{s_r}}\! \}}{\CMcal{S}_{01}}\left[{1 - \Gamma _2^{\{\! {{s_r}}\! \}}\left( {{\CMcal{S}_{11}} - {\CMcal{S}_{21}}} \right)}\right]}{{1\!-\!{\CMcal{S}_{11}}\!\!\left( {\Gamma _1^{\{\! {{s_r}}\! \}}\!\!+\! \Gamma _2^{\{\! {{s_r}}\! \}}}\! \right)\!\! + \!\Gamma _1^{\{\! {{s_r}}\! \}}\Gamma _2^{\{\! {{s_r}}\! \}}\!\left( \!{\CMcal{S}_{11}^2 \!\!-\! \CMcal{S}_{21}^2} \right)}}}\\
{\ell _2^{\{\! {Z_1^{\{\! {{s_r}}\! \}},Z_2^{\{\! {{s_r}}\! \}}}\! \}}\!\! = \!\frac{\Gamma _2^{\{\! {{s_r}}\! \}}{\CMcal{S}_{01}}\left[{1 - \Gamma _1^{\{\! {{s_r}}\! \}}\left( {{\CMcal{S}_{11}} - {\CMcal{S}_{21}}} \right)}\right]}{{1\!-\!{\CMcal{S}_{11}}\!\!\left( {\Gamma _1^{\{\! {{s_r}}\! \}}\!\!+\! \Gamma _2^{\{\! {{s_r}}\! \}}}\! \right)\!\! + \!\Gamma _1^{\{\! {{s_r}}\! \}}\Gamma _2^{\{\! {{s_r}}\! \}}\!\left( \!{\CMcal{S}_{11}^2 \!\!-\! \CMcal{S}_{21}^2} \right)}} \cdot }
\end{align}
\end{subequations}
After some mathematical manipulations on (\ref{eq:eq37}), (\ref{eq:eq38}) and (\ref{eq:eq25}), the loads $Z_1^{\left\{ {{s_r}} \right\}}$ and $Z_2^{\left\{ {{s_r}} \right\}}$ can directly be found by solving the following equations,
\begin{subequations}\label{eq:eq39}
\begin{align}
{\Gamma _1^{\{\! {{s_r}}\! \}} \!\!=\! \frac{{\Gamma_\text{II}}\left({1\!\!+\! {s_r}}\right)\! +\! {\Gamma_\text{I}}\left({1\!\! -\! {s_r}}\right)\! -\! 2{\Gamma_\text{I}}{\Gamma_\text{II}}\!\left( {{\CMcal{S}_{11}}\!\! -\! {\CMcal{S}_{21}}} \right)}{{2 - \left[ {{\Gamma_\text{I}}\left({1 + {s_r}}\right) + {\Gamma_\text{II}}\left({1 - {s_r}}\right)} \right]\left( {{\CMcal{S}_{11}} - {\CMcal{S}_{21}}} \right)}}}\\[2ex]
{\Gamma _2^{\{\! {{s_r}}\! \}} \!\!=\! \frac{{\Gamma_\text{II}}\left({1\!\!-\! {s_r}}\right)\! +\! {\Gamma_\text{I}}\left({1\!\! +\! {s_r}}\right)\! -\! 2{\Gamma_\text{I}}{\Gamma_\text{II}}\!\left( {{\CMcal{S}_{11}}\!\! -\! {\CMcal{S}_{21}}} \right)}{{2 - \left[ {{\Gamma_\text{I}}\left({1 - {s_r}}\right) + {\Gamma_\text{II}}\left({1 + {s_r}}\right)} \right]\left( {{\CMcal{S}_{11}} - {\CMcal{S}_{21}}} \right)}}}.
\end{align}
\end{subequations}

\subsection{Discussion and Implementation}\label{section2e}
\vspace{-1em}
Equation (\ref{eq:eq39}) shows that the control load pair required for single-radio multiplexing of the symbol pair $\{s_1{,}s_2\}$ depend on their symbol combination ratio $s_r$. This demands a distinct load pair at the control ports for each possible symbol combination ratio of the considered modulation. For instance, in the case of an \emph{M}-PSK modulation scheme, since there are \emph{M} different values of $s_r$, \emph{M} distinct load values are required at each control port for enabling the proposed single-radio multiplexing. On the other hand, it is seen from (\ref{eq:eq39}) that altering the polarity of the combination ratio (i.e. ${s_r} \to  - {s_r}$) swaps the loads at the control ports (i.e. $\Gamma _1^{\left\{ {{s_r}} \right\}} \leftrightarrow \Gamma _2^{\left\{ {{s_r}} \right\}}$). This implies that in the case of rotationally symmetric constellations exactly the same set of load values is required at both control ports.

In the special case of BPSK signaling, the symbol combination ratio is either $+1$ or $-1$, i.e. ${s_r} =  \pm 1$. Using (\ref{eq:eq39}) it can be seen that the control load values $Z_1^{\left\{ { \pm 1} \right\}}$ and $Z_2^{\left\{ { \pm 1} \right\}}$ are identical to the basis impedances (i.e. the ones used when defining the basis functions),
\begin{subequations}\label{eq:eq40}
\begin{align}
\Gamma _1^{\left\{ { + 1} \right\}} = \Gamma _2^{\left\{ { - 1} \right\}} = {\Gamma_\text{II}}\\
\Gamma _2^{\left\{ { + 1} \right\}} = \Gamma _1^{\left\{ { - 1} \right\}} = {\Gamma_\text{I}} \cdot
\end{align}
\end{subequations}
This is in full agreement with the results from the earlier work \cite{alrabadi_tap} (where a technique limited to the BPSK signaling cases was presented) and demonstrates the validity of our proposed approach at least while dealing with BPSK modulation scheme.

As mentioned in the introduction, one of the main practical limitations of previous art is related to the large dynamic variation of the antenna system input impedance associated with the control loads reconfiguration. By contrast, the technique proposed here provides a constant impedance matching for all possible symbol combinations of $s_1$ and $s_2$. Using (\ref{eq:eq14}), (\ref{eq:eq37}) and (\ref{eq:eq25}), the total reflection coefficient at the active port becomes
\begin{align}\label{eq:eq41}
\!\!\!\Gamma _\text{tot}^{\{\! {Z_1^{\{\! {{s_r}}\! \}},Z_2^{\{\! {{s_r}}\! \}}}\! \}}\!\! &=\! {\CMcal{S}_{00}} \!+\! {\CMcal{S}_{01}}\left[ {\ell _1^{\{\! {Z_1^{\{\! {{s_r}}\! \}},Z_2^{\{\! {{s_r}}\! \}}}\! \}} + \ell _2^{\{\! {Z_1^{\{\! {{s_r}}\! \}},Z_2^{\{\! {{s_r}}\! \}}}\! \}}} \right] \nonumber\\
&=\! {\CMcal{S}_{00}}\! + \!{\CMcal{S}_{01}}\left[ {2\ell _{{\CMcal{B}_1}}^{\left\{ {{Z_\text{I}},{Z_\text{II}}} \right\}}} \right]\\
&=\! {\CMcal{S}_{00}} \!+ \!{\CMcal{S}_{01}}\left[ {\ell _1^{\left\{ {{Z_\text{I}},{Z_\text{II}}} \right\}} + \ell _2^{\left\{ {{Z_\text{I}},{Z_\text{II}}} \right\}}} \right]\nonumber\\
&=\!\CMcal{S}_{00}\!+\!\frac{\CMcal{S}_{01}^2\left[{{\Gamma_\text{I}} + {\Gamma_\text{II}} - 2{\Gamma_\text{I}}{\Gamma_\text{II}}\left( {{\CMcal{S}_{11}} \!-\! {\CMcal{S}_{21}}} \right)}\right]}{{1\!-\! {\CMcal{S}_{11}}\left( {{\Gamma_\text{I}}\! +\! {\Gamma_\text{II}}} \right)\! +\! {\Gamma_\text{I}}{\Gamma_\text{II}}\left( {\CMcal{S}_{11}^2\!-\!\CMcal{S}_{21}^2} \right)}} \nonumber
\end{align}
which remains constant regardless of the symbol combination ratio $s_r$. This is of great practical importance as no external reconfigurable matching network at the active port is required.

\begin{figure}[t!]
\centering
\includegraphics[width=4.2in]{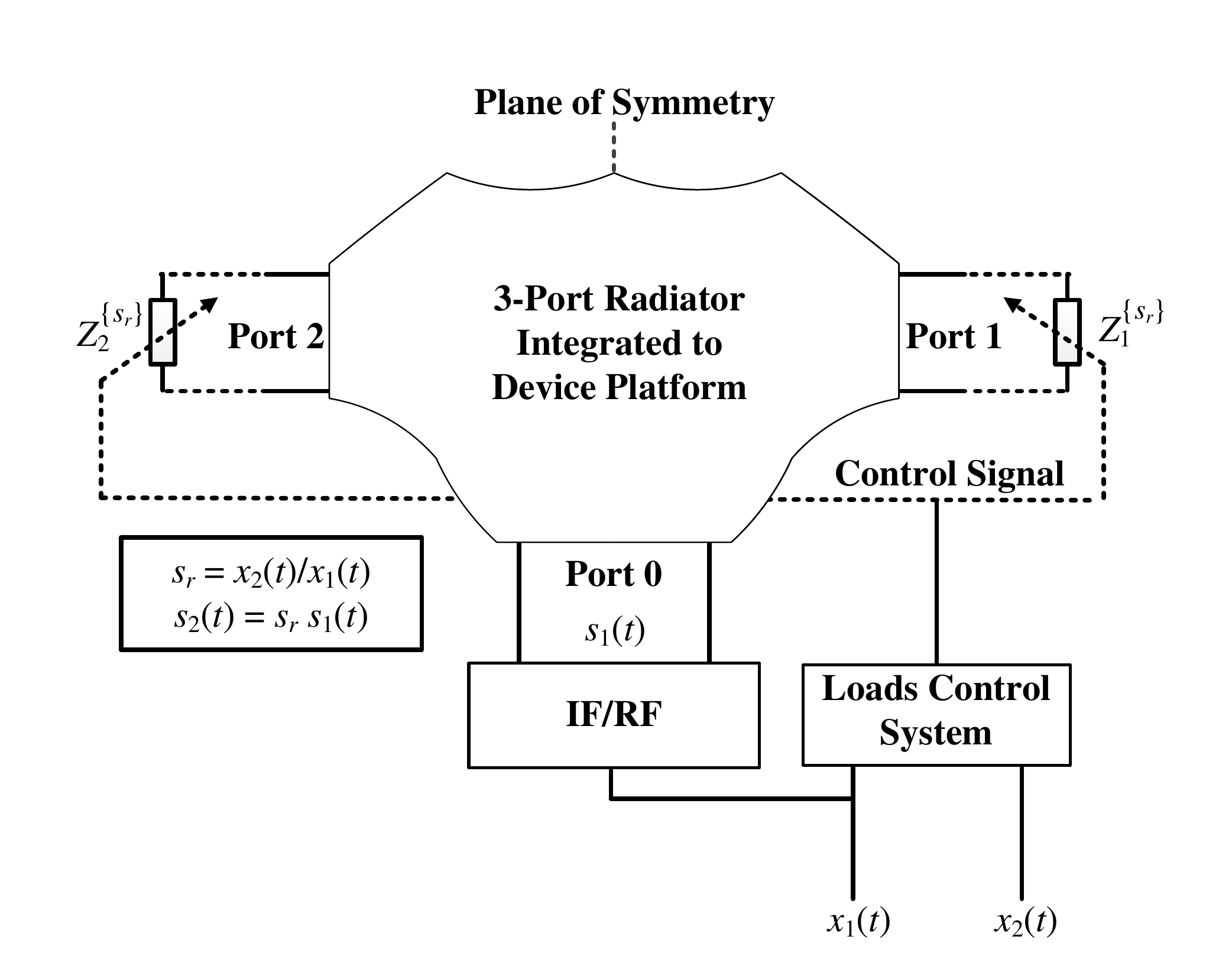}
\caption{Symbolic representation of the proposed system capable of multiplexing two input symbol streams. $x_1(t)$ and $x_2(t)$ are the data streams in the baseband domain. IF: intermediate frequency.}
\label{fig:fig3}
\end{figure}

We have so far shown that a reconfigurable antenna composed of a symmetric three-port radiator and two variable loads is capable of transmitting two symbol streams of any modulation scheme. Fig. \ref{fig:fig3} depicts an antenna system solution based on the proposed approach. The inputs to the system consist of two streams of symbols in the baseband domain $x_1(t)$ and $x_2(t)$. The first stream $x_1(t)$ is upconverted to $s_1(t)$ and fed into the antenna central active port. Unlike the classical MIMO, the second stream $x_2(t)$ does not leave the digital signal processing (DSP) unit. A loads control system provides the control signal for reconfiguring the variable loads at the control ports according to the ratio of two symbols in the baseband domain $x_1(t)$ and $x_2(t)$. By doing this, two data streams, i.e. the real $s_1(t)$ and the virtual ${s_{\rm{2}}}\left( t \right) = {{{s_{\rm{1}}}\left( t \right){x_{\rm{2}}}\left( t \right)} \mathord{\left/
 {\vphantom {{{s_{\rm{1}}}\left( t \right){x_{\rm{2}}}\left( t \right)} {{x_{\rm{1}}}\left( t \right)}}} \right.
 \kern-\nulldelimiterspace} {{x_{\rm{1}}}\left( t \right)}}$, are independently mapped onto each basis pattern in the beam-space domain.

\subsection{Passive Loading Constraint}\label{section2f}
For the sake of completeness, all the derivations so far considered the utilization of complex-valued impedances at the control ports. However, the use of a load with non-negligible positive real part degrades the radiation efficiency of the antenna system, while employing active loads would drastically increase the implementation complexity and also potentially lead to stability issues. In this context, it is desirable to analyze the proposed approach when constrained to only purely imaginary load solutions which are more attractive for realistic applications.

The condition of purely imaginary loads at the control ports (i.e. $Z_1^{\left\{ {{s_r}} \right\}} = jX_1^{\left\{ {{s_r}} \right\}}$ and $Z_2^{\left\{ {{s_r}} \right\}} = jX_2^{\left\{ {{s_r}} \right\}}$) is equivalent to
\begin{equation}\label{eq:eq42}
\left| {\Gamma _1^{\left\{ {{s_r}} \right\}}} \right| = \left| {\Gamma _2^{\left\{ {{s_r}} \right\}}} \right| = 1
\end{equation}
for all possible symbol combination ratios $s_r$. This implies that according to (\ref{eq:eq40}) the basis impedances are also purely imaginary, i.e. $Z_\text{I} = jX_\text{I}$ and $Z_\text{II} = jX_\text{II}$. Applying (\ref{eq:eq42}) to (\ref{eq:eq39}), ${\Gamma_\text{I}}$ and ${\Gamma_\text{II}}$ must simultaneously satisfy two following equations:
\begin{subequations}\label{eq:eq43}
\begin{align}
&\frac{{1 - {{\left| {\Delta} \right|}^2}}}{{1 + {{\left| {\Delta} \right|}^2}}}  \frac{{1 - {{\left| {{s_r}} \right|}^2}}}{{2\left| {{s_r}} \right|}}\frac{\sin \left[ \frac{1}{2}(\vartheta _{\Gamma_\text{I}} - \vartheta _{\Gamma_\text{II}})\right]}{\sin {\vartheta _{{s_r}}}} = \\
&\cos \left[ \tfrac{1}{2}(\vartheta _{\Gamma_\text{I}} - \vartheta _{\Gamma_\text{II}}) \right] - \frac{{2\left| {\Delta} \right| }}{{1 + {{\left| {\Delta} \right|}^2}}}\cos \left[ \vartheta _{\Delta } + \tfrac{1}{2}(\vartheta _{\Gamma_\text{I}} + \vartheta _{\Gamma_\text{II}}) \right]  \nonumber \\[2ex]
&\frac{{1 - {{\left| {\Delta} \right|}^2}}}{{1 + {{\left| {\Delta} \right|}^2}}}  \frac{{1 - {{\left| {{s_r}} \right|}^2}}}{{2\left| {{s_r}} \right|}}\frac{\sin \left[ \frac{1}{2}(\vartheta _{\Gamma_\text{I}} - \vartheta _{\Gamma_\text{II}})\right]}{-\sin {\vartheta _{{s_r}}}} = \\
&\cos \left[ \tfrac{1}{2}(\vartheta _{\Gamma_\text{I}} - \vartheta _{\Gamma_\text{II}}) \right] - \frac{{2\left| {\Delta} \right| }}{{1 + {{\left| {\Delta} \right|}^2}}}\cos \left[ \vartheta _{\Delta } + \tfrac{1}{2}(\vartheta _{\Gamma_\text{I}} + \vartheta _{\Gamma_\text{II}}) \right]  \nonumber
\end{align}
\end{subequations}
where for the sake of compactness, we denoted ${\Gamma_\text{I}} = \exp (j{\vartheta _{{\Gamma_\text{I}}}})$,  ${\Gamma_\text{II}} = \exp (j{\vartheta _{{\Gamma_\text{II}}}})$,  ${s_r} = \left| {{s_r}} \right|\exp \left( {j{\vartheta _{{s_r}}}} \right)$ and  ${\CMcal{S}_{11}} - {\CMcal{S}_{21}} = \left| {\Delta } \right|\exp \left( {j{\vartheta _{\Delta }}} \right)$. In general, such a solution for ${\Gamma_\text{I}}$ and ${\Gamma_\text{II}}$ does not exist. However, for the particular case $\left| {{s_r}} \right| = {\rm{1}}$, namely for PSK modulation, both equations in (\ref{eq:eq43}) become equivalent:
\begin{equation}\label{eq:eq44}
\cos \! \left[ \tfrac{1}{2}(\vartheta _{\Gamma_\text{I}}\!\! -\! \vartheta _{\Gamma_\text{II}}) \right]\!\!-\!\! \frac{{2\left| {\Delta} \right| }}{{1 \!\!+\! {{\left| {\Delta} \right|}^2}}}\cos \left[ \vartheta _{\Delta } \!+ \! \tfrac{1}{2}(\vartheta _{\Gamma_\text{I}} \!\!+\! \vartheta _{\Gamma_\text{II}}) \right] \!\!=\! 0 .
\end{equation}

This equation provides a bijective mapping between the basis reactances $X_\text{I}$ and $X_\text{II}$. In other words, each imaginary impedance $jX_\text{I}$ is paired with a unique imaginary impedance $jX_\text{II}$. Accordingly, other reactances $X_1^{\left\{ {{s_r}} \right\}}$ and $X_2^{\left\{ {{s_r}} \right\}}$ for the considered PSK modulation scheme are calculated using (\ref{eq:eq39}). These results are of significant practical importance: the proposed technique still allows the single-radio multiplexing of higher order PSK data streams with a single reconfigurable antenna when (i) only the use of purely reactive loads is permitted, and (ii) no reconfigurable impedance matching circuit is utilized.

Applying the reactive load condition in (\ref{eq:eq44}) to (\ref{eq:eq41}), the total reflection coefficient at the active port can be expressed in terms of the scattering parameters only,
\begin{align}\label{eq:neweq45}
\Gamma _\text{tot}^{\{\! {Z_1^{\{\! {{s_r}}\! \}},Z_2^{\{\! {{s_r}}\! \}}}\! \}}\!\! =\! \CMcal{S}_{00}\! + \!\CMcal{S}_{01}^2\frac{{2( \CMcal{S}_{11} - \CMcal{S}_{21} )}^*}{{1\! - \!{{( {{{\CMcal{S}}_{11}}\! -\! {{\CMcal{S}}_{21}}} )}^*}( {{{\CMcal{S}}_{11}} \!+\! {{\CMcal{S}}_{21}}} )}}.
\end{align}
Similarly, it can be demonstrated that satisfying the passive loading constraint also removes the dependency of the basis power imbalance ratio on the basis loads (see Appendix \ref{appendix:I}). Therefore, having the reactance $X_\text{I}$ as a free parameter, we can optimize the antenna system according to a specific criterion in terms of the basis power imbalance ratio and the total efficiency.

\section{Design Procedure and Antenna Example} \label{section3}
In this section, we illustrate the proposed approach by designing a compact antenna system which is capable of transmitting two QPSK data streams using a single RF chain. The antenna is designed on the small platform of a hypothetical USB dongle, modeled by a 1.6-mm-thick FR4 substrate of 20 mm $\times$ 45 mm with a dielectric constant of 4.4. The substrate area is very small, about 0.0625${\lambda ^{\rm{2}}}$ at the design frequency of 2.5 GHz. In the following, the step-by-step design procedure is fully described and the simulation results demonstrating the efficiency of the approach are presented.

\begin{figure}[t!]
\centering
\includegraphics[width=4.1in]{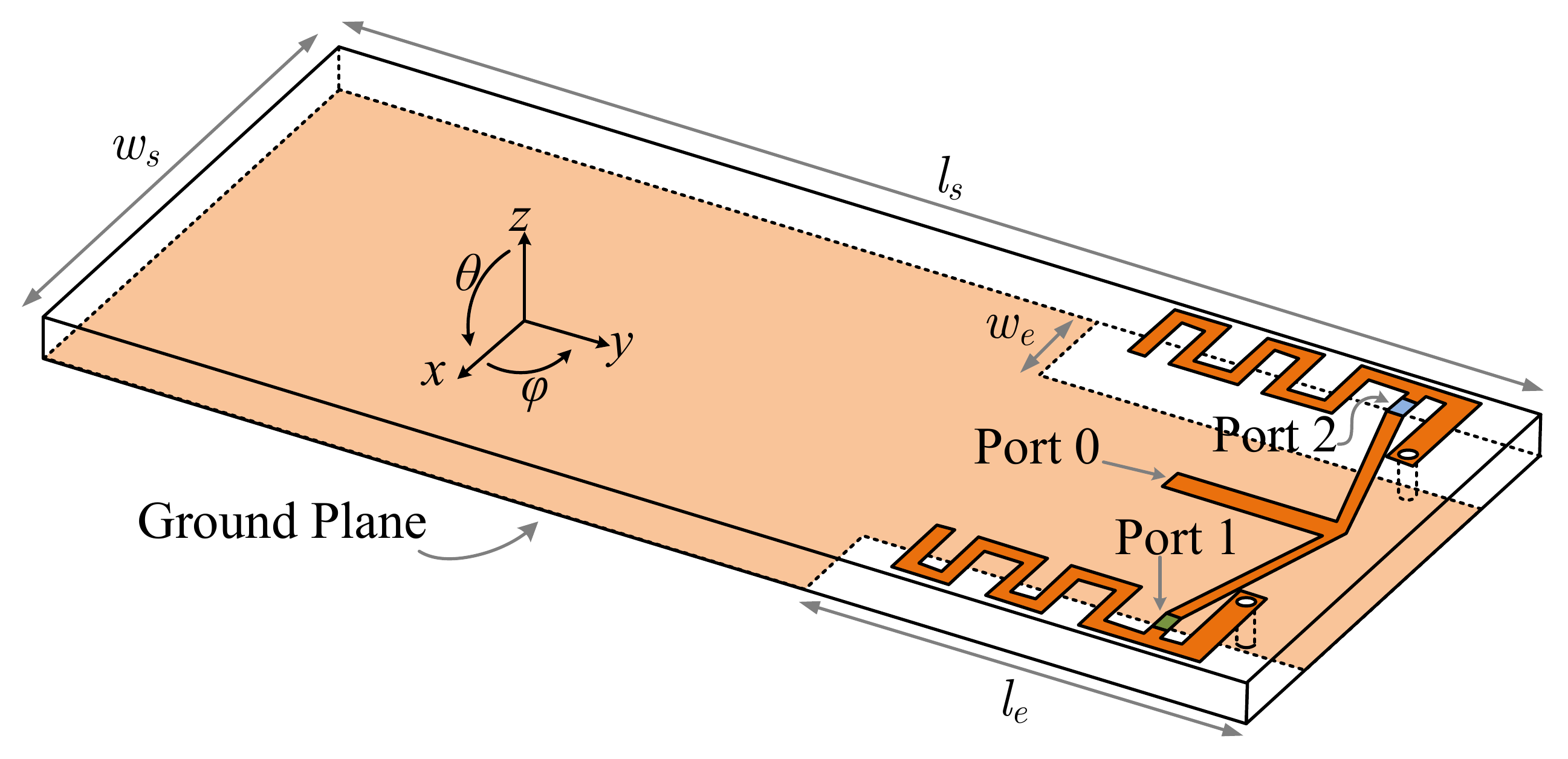}
\caption{Designed compact symmetric three-port radiator. $l_s =$ 45 mm, $w_s =$ 20 mm, $l_e =$ 16.5 mm, $ w_e =$ 4 mm.}
\label{fig:fig4}
\end{figure}

As a first step, a symmetric three-port radiator such as the one depicted in Fig. \ref{fig:fig3} should be designed according to the physical requirements of the desired application. Fig. \ref{fig:fig4} shows the designed radiating structure, printed on the FR4 substrate, with an axis of symmetry in \emph{yz}-plane. The central port is considered as the only active port for the connection to the single RF module. The two lateral ports are the control ones, which will be terminated with purely imaginary reconfigurable loads whose values are calculated using the developed formulation.

Table \ref{table:table1} shows the possible symbol combination ratios of two QPSK signals, where ${[{b_1}{b_2}\;{b_3}{b_4}]^\text{T}}$ is the input vector of bits modulated into ${[{x_1}\;{x_2}]^\text{T}}$. There are four distinct symbol combination ratios, thus four system states are sufficient for transmitting the two QPSK data streams. In each system state, the control ports are terminated with two distinct load values. However, since QPSK is a rotationally symmetric modulation scheme (see Section \ref{section2e}), the same set of four reactance values can be used at each port: the first two reactance values associated with the States 1 and 2 are identical to the basis reactances $X_\text{I}$ and $X_\text{II}$, and the other two related to the States 3 and 4 are obtained using (\ref{eq:eq39}) when $s_r = \pm j$.

\begin{figure}[t!]
\centering
\includegraphics[width=4.5in]{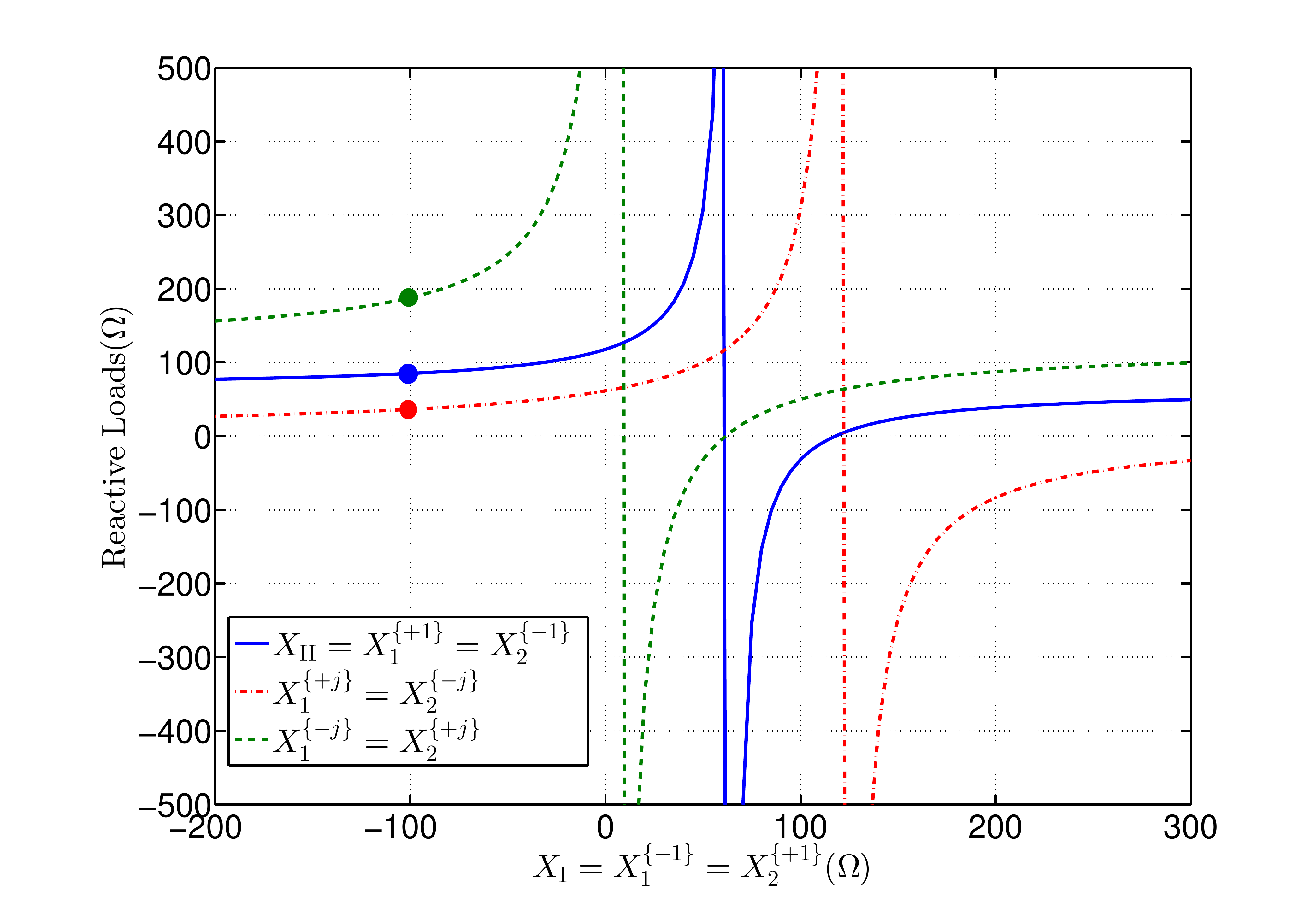}
\caption{Required reactive loads as a function of $X_\text{I}$ for QPSK beam-space multiplexing.}
\label{fig:fig5}
\vspace{-1em}
\end{figure}

The scattering parameters and the embedded radiated fields of the three-port system are extracted from electromagnetic full-wave simulation, here using Ansys HFSS. The resulting scattering matrix at the design frequency is given by
\begin{align*}
\boldsymbol{\CMcal{S}} = \left[ {\begin{array}{*{3}{c}}
{0.24 + j0.19}&{ - 0.13 + j0.47}&{ - 0.13 + j0.47}\\
{ - 0.13 + j0.47}&{0.46 - j0.27}&{0.14 + j0.13}\\
{ - 0.13 + j0.47}&{0.14 + j0.13}&{0.46 - j0.27}
\end{array}} \right]
\end{align*}
and used to find the pairs of the basis reactances $X_\text{I}$ and $X_\text{II}$ using (\ref{eq:eq44}), ensuring that the reactive load condition in (\ref{eq:eq42}) is satisfied. Then, for each combination ratio the required control impedances are obtained using (\ref{eq:eq39}). As stated in Section \ref{section2f}, $X_\text{I}$ is a free parameter of the antenna system. Therefore, as shown in Fig. \ref{fig:fig5}, the other three reactive loads required for QPSK signaling can be represented as a function of $X_\text{I}$. The optimal set of reactive impedances might be selected regarding the availability and the practical realization of the reactive loads.
For any arbitrary set of the obtained reactances, the total reflection coefficient at the active port can be calculated using (\ref{eq:neweq45}). A return loss of 10.4 dB is achieved thanks to the optimization of the three-port radiator.

\begin{table}[!t]
\caption{Two QPSK Symbols Combinations}
\renewcommand{\arraystretch}{0.8}
\centering
\footnotesize
\begin{tabular}{c c c c}
\hline \hline
\\[-1.5ex]
${[{b_1}{b_2}\;{b_3}{b_4}]^\text{T}}$ &
${[{x_1}\;{x_2}]^\text{T}}$& {$s_r$}  & {System State} \\ [0.5ex]
\hline \\[-1.5ex]
${[{00}\;{00}]^\text{T}}$ &
${[{e^{-\tfrac{j3\pi}{4}}}\;{e^{-\tfrac{j3\pi}{4}}}]^\text{T}}$
& $+1$ & 2\\
${[{00}\;{01}]^\text{T}}$ &
${[{e^{-\tfrac{j3\pi}{4}}}\;{e^{+\tfrac{j3\pi}{4}}}]^\text{T}}$
& $-j$ & 4\\
${[{00}\;{11}]^\text{T}}$ &
${[{e^{-\tfrac{j3\pi}{4}}}\;{e^{+\tfrac{j\pi}{4}}}]^\text{T}}$
& $-1$ & 1\\
${[{00}\;{10}]^\text{T}}$ &
${[{e^{-\tfrac{j3\pi}{4}}}\;{e^{-\tfrac{j\pi}{4}}}]^\text{T}}$
 & $+j$ & 3\\
\hline\\[-1.5ex]
${[{01}\;{00}]^\text{T}}$ &
${[{e^{+\tfrac{j3\pi}{4}}}\;{e^{-\tfrac{j3\pi}{4}}}]^\text{T}}$
& $+j$ & 3\\
${[{01}\;{01}]^\text{T}}$ &
${[{e^{+\tfrac{j3\pi}{4}}}\;{e^{+\tfrac{j3\pi}{4}}}]^\text{T}}$
& $+1$ & 2\\
${[{01}\;{11}]^\text{T}}$ &
${[{e^{+\tfrac{j3\pi}{4}}}\;{e^{+\tfrac{j\pi}{4}}}]^\text{T}}$
& $-j$ & 4\\
${[{01}\;{10}]^\text{T}}$ &
${[{e^{+\tfrac{j3\pi}{4}}}\;{e^{-\tfrac{j\pi}{4}}}]^\text{T}}$
 & $-1$ & 1\\
\hline\\[-1.5ex]
${[{11}\;{00}]^\text{T}}$ &
${[{e^{+\tfrac{j\pi}{4}}}\;{e^{-\tfrac{j3\pi}{4}}}]^\text{T}}$
& $-1$ & 1\\
${[{11}\;{01}]^\text{T}}$ &
${[{e^{+\tfrac{j\pi}{4}}}\;{e^{+\tfrac{j3\pi}{4}}}]^\text{T}}$
& $+j$ & 3\\
${[{11}\;{11}]^\text{T}}$ &
${[{e^{+\tfrac{j\pi}{4}}}\;{e^{+\tfrac{j\pi}{4}}}]^\text{T}}$
& $+1$ & 2\\
${[{11}\;{10}]^\text{T}}$ &
${[{e^{+\tfrac{j\pi}{4}}}\;{e^{-\tfrac{j\pi}{4}}}]^\text{T}}$
 & $-j$ & 4\\
\hline\\[-1.5ex]
${[{10}\;{00}]^\text{T}}$ &
${[{e^{-\tfrac{j\pi}{4}}}\;{e^{-\tfrac{j3\pi}{4}}}]^\text{T}}$
& $-j$ & 4\\
${[{10}\;{01}]^\text{T}}$ &
${[{e^{-\tfrac{j\pi}{4}}}\;{e^{+\tfrac{j3\pi}{4}}}]^\text{T}}$
& $-1$ & 1\\
${[{10}\;{11}]^\text{T}}$ &
${[{e^{-\tfrac{j\pi}{4}}}\;{e^{+\tfrac{j\pi}{4}}}]^\text{T}}$
& $+j$ & 3\\
${[{10}\;{10}]^\text{T}}$ &
${[{e^{-\tfrac{j\pi}{4}}}\;{e^{-\tfrac{j\pi}{4}}}]^\text{T}}$
 & $+1$ & 2\\
\hline
\end{tabular} \label{table:table1}
\end{table}

%

The antenna embedded radiated fields extracted from the full-wave simulation are then used for calculating the power radiated by the basis patterns using (\ref{eq:eqnew47}). A power imbalance ratio of 1.04 between the basis patterns is obtained, showing that the designed antenna system is capable of creating nearly balanced basis patterns due to the optimization of the radiator. It is worth noting that far-field calculations of the power imbalance ratio for the designed antenna were carried out here as the utilized substrate is quite lossy (with a loss tangent of 0.02) and the antenna is of compact dimensions. However, when the loss in the radiator materials is negligible, the power imbalance ratio can be directly calculated in terms of the radiator scattering parameters (see (\ref{eq:eq46}) and (\ref{eq:eqnew46}) in Appendix \ref{appendix:I}).

\begin{figure}[t!]
\centering
\includegraphics[width=4.5in]{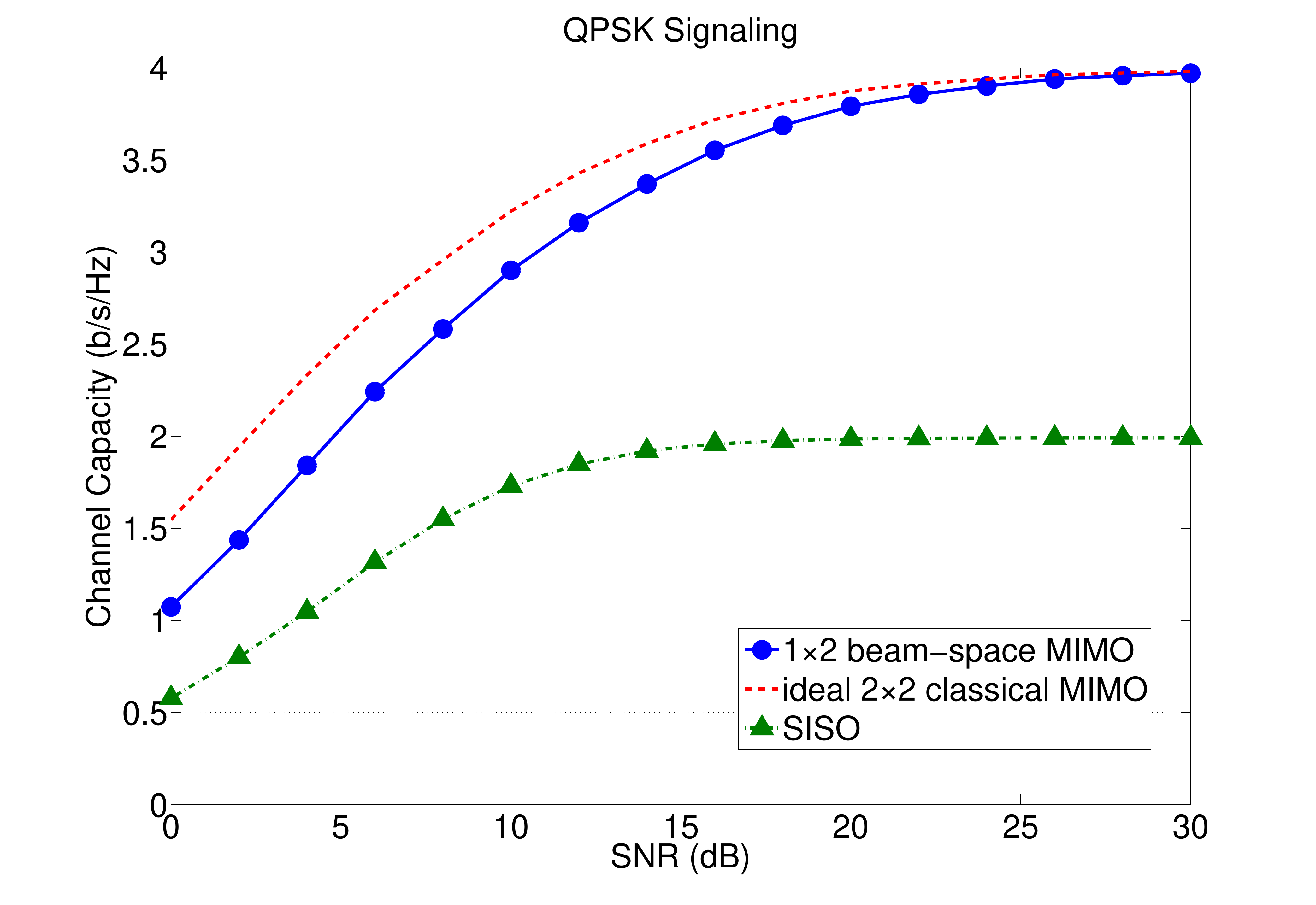}
\caption{Channel capacity of the antenna system as a function of the transmit SNR under QPSK signaling when $X_\text{I} = - 100\;\Omega$. The simulation was performed based on Monte-Carlo method (see \cite{montecarlo}) for an input of $10000$ uniformly distributed QPSK symbols where $10000$ channel realizations were used. Pulse shaping is not included.}
\label{fig:fig7}
\vspace{-1em}
\end{figure}

To provide some insight into the multiplexing performance of the designed antenna system, we computed the system capacity under QPSK signaling based on the full-wave simulation results for an arbitrary set of reactance values  (i.e. the set associated with $X_\text{I} = - 100\;\Omega$). Two QPSK signals are simultaneously transmitted over two orthogonal basis patterns while assuming a Kronecker narrowband flat-fading channel \cite{Kronecker}. The transmitted signals then received using two uncorrelated and uncoupled antenna elements in an open-loop MIMO operation. Thus the channel transfer function can be written as
\begin{equation}\label{eq:eq48}
{{\textbf{H}}_{{\text{ch}}}} = {{\textbf{H}}_{\text{w}}}{\textbf{R}}_{\text{T}}^{{1 \mathord{\left/
 {\vphantom {1 2}} \right.
 \kern-\nulldelimiterspace} 2}}
\end{equation}
where the elements of the matrix ${{\textbf{H}}_{\text{w}}} \in {\mathbb{C}^{2 \times 2}}$ are independent and identically distributed (i.i.d) complex Gaussian random variables with zero mean and unit variance. Since the basis patterns are defined when exciting the active port with a unit power, the transmit covariance matrix $\textbf{R}_{\text{T}}$ is obtained as
\begin{equation}\label{eq:eq49}
{\textbf{R}}_{\rm{T}} = {P_{{\text{in}}}}\left[ {\begin{array}{*{20}{c}}
{{{\CMcal P}_{{{\CMcal B}_1}}}}&{{\chi _{{{\CMcal B}_1}{{\CMcal B}_2}}}}\\
{{\chi _{{{\CMcal B}_2}{{\CMcal B}_1}}}}&{{{\CMcal P}_{{{\CMcal B}_2}}}}
\end{array}} \right] = {P_{{\text{in}}}}\,{\text{diag}}\left[ {{{\CMcal P}_{{{\CMcal B}_1}}},{{\CMcal P}_{{{\CMcal B}_2}}}} \right]
\end{equation}
where $P_{{\text{in}}}$ is the input power. As shown in Fig. \ref{fig:fig7}, the channel capacity of the designed beam-space MIMO system in higher signal to noise ratio (SNR) converges to that of an ideal 2$\times$2 classical MIMO system having an identity transmit correlation matrix. The discrepancy between the curves in the low-SNR region is due to the total efficiency of the designed antenna. A total efficiency of $56\%$ at the working frequency is obtained which is mainly due to the dielectric loss in the substrate. For the sake of comparison, Fig. \ref{fig:fig7} also shows the capacity curve of a single-input single-output (SISO) system having the same transmit total efficiency. We can thus conclude that proposed multiplexing approach performs as expected.

According to Table \ref{table:table1}, the ratio of the second and first symbol streams $s_r$ determines the system state and consequently the states of the control loads. When the ratio $s_r$ remains constant during the symbol transition, the states of the control loads are not altered and no pattern reconfigurability occurs. As the transmission concept includes a pulse shaping filter in the path of the first datastream $s_1$, i.e. the one directly fed to the single RF chain, the beam-space MIMO system can practically fulfill required spectral mask constraints. On the other side, during other symbol transitions (i.e. when $s_r$ changes), the control signal switches the states of the control loads and the antenna instantaneous radiation pattern is altered. In such symbol transitions, improper transition between the states of the control loads may give rise to bandwidth expansion of transmitted signals \cite{my_tap}. One potential solution might lie in controlling the transition among different states of the control loads. A more detailed study on this issue is an important topic of further investigation, but is out of the scope of the present contribution.

\section{Conclusion}
\vspace{-0em}
An efficient approach for multiplexing two PSK data streams of any modulation order via a single RF chain and a single reconfigurable antenna has been described. The approach not only provides a constant impedance response over all the operational states, but also uses purely imaginary loads at the control ports. Moreover, it replaces the multiplexing of basis patterns with the multiplexing of basis vectors. The theory and design method were successfully illustrated by the first example of a realistic compact single-radio antenna capable of transmitting two QPSK data streams with passive loads and constant input impedance. These results constitute a crucial step towards MIMO with simpler and cheaper RF hardware for real-life wireless terminals. Future work in this very promising field should be directed towards the issue of the out-of-band radiation associated with transient behavior of the embedded reconfigurable elements.

\clearpage


%

\appendices

\section{Proof of (\ref{eq:eq10})}\label{appendix:II}
Let first rewrite (\ref{eq:eq9}) as
\begin{equation}\label{eq:eqap1}
\boldsymbol{a}^\text{H}{\textbf{U}}\boldsymbol{a} = 0
\end{equation}
where ${\textbf{U}} = \boldsymbol{\CMcal X} + \boldsymbol{\CMcal S}^\text{H}\boldsymbol{\CMcal S} - {\textbf{I}}$ is an $N$-by-$N$ matrix. Therefore, we obtain that
\begin{equation}\label{eq:eqap2}
\sum\limits_{m = 0}^{N - 1} {a_m^{\rm{*}}{{\left( {{\textbf{U}}\boldsymbol{a}} \right)}_m}}  = \sum\limits_{m = 0}^{N - 1} {\sum\limits_{n = 0}^{N - 1} {a_m^{\rm{*}}{U_{mn}}{a_n}} }  = 0.
\end{equation}
Let  ${a_m} = {\delta _{rm}}$, which is 1 only for index $r$ and zero elsewhere. Then,
\begin{equation}\label{eq:eqap3}
\boldsymbol{a}^\text{H}{\textbf{U}}\boldsymbol{a} = U_{rr} = 0.
\end{equation}
Therefore, the diagonal elements of the matrix \textbf{U} are zero.

On the other hand, letting  ${a_m} = {\delta _{sm}} + {\delta _{tm}}$, we have
\begin{equation}\label{eq:eqap4}
\boldsymbol{a}^\text{H}{\textbf{U}}\boldsymbol{a} = U_{st} + U_{ts} = 0.
\end{equation}
Therefore, the matrix \textbf{U} is anti-symmetric. However, letting  ${a_m} = j{\delta _{sm}} + {\delta _{tm}}$, we obtain that
\begin{equation}\label{eq:eqap5}
\boldsymbol{a}^\text{H}{\textbf{U}}\boldsymbol{a} = -jU_{st} + jU_{ts} = 0.
\end{equation}
Thus the off-diagonal elements are anti-symmetric and equal, hence zero.

\section{Calculation of Power Radiated by Basis Patterns}\label{appendix:I}
The total power radiated in the far-field by the basis patterns defined in (\ref{eq:eq23}) can be obtained using (\ref{eq:eq8}),
\begin{subequations}\label{eq:eq45}
\begin{align}
\CMcal{P}_{{\CMcal{B}_1}} &= \chi _{{\CMcal{B}_1}\!{\CMcal{B}_1}} =\! \frac{{{\eta _0}}}{2}\!\!\int\!\!\!\!\!\int\!\! {{{\left| {{\CMcal{B}_1( {\theta{,}\varphi })}} \right|}^2}dS} \! = \!\frac{{{\eta _0}}}{2}\!\!\int\!\!\!\!\!\int \!{{{\left| {\boldsymbol{v}_{{\CMcal{B}_1}}^{\{ {{Z_\text{I}},{Z_\text{II}}} \}}\boldsymbol{\CMcal{E}}_\text{emb}^T} \right|}^2}\!dS} \nonumber \\
&= \!\frac{{{\eta _0}}}{2}\!\!\int\!\!\!\!\!\int\!\! {{{\left| {{\CMcal{E}_0( {\theta{,}\varphi })} + \ell _{{\CMcal{B}_1}}^{\{ {{Z_\text{I}},{Z_\text{II}}} \}}\left[ {{\CMcal{E}_1( {\theta{,}\varphi })} + {\CMcal{E}_2}( {\theta{,}\varphi })} \right]} \right|}^2}dS} \nonumber\\
&= {\CMcal{P}_{{\CMcal{E}_0}}} + {\left| {\ell _{{\CMcal{B}_1}}^{\{ {{Z_\text{I}},{Z_\text{II}}} \}}} \right|^2}\left( {{\CMcal{P}_{{\CMcal{E}_1}}} + {\CMcal{P}_{{\CMcal{E}_2}}} + {\chi _{12}} + {\chi _{21}}} \right)\nonumber\\
&\quad + \ell _{{\CMcal{B}_1}}^{\{ {{Z_\text{I}},{Z_\text{II}}} \}}\left( {{\chi _{01}} + {\chi _{02}}} \right) + {\ell {_{{\CMcal{B}_1}}^{\{ {{Z_\text{I}},{Z_\text{II}}} \}}}}^*\left( {{\chi _{10}} + {\chi _{20}}} \right)\nonumber\\
& = {\CMcal{P}_{{\CMcal{E}_0}}} + 2{\left| {\ell _{{\CMcal{B}_1}}^{\{ {{Z_\text{I}},{Z_\text{II}}} \}}} \right|^2}\left( {{\CMcal{P}_{{\CMcal{E}_1}}} + {\chi _{21}}} \right) \nonumber\\
&\quad + 4{\mathop{\rm Re}\nolimits} \left\{ {\ell _{{\CMcal{B}_1}}^{\{ {{Z_\text{I}},{Z_\text{II}}} \}}{\chi _{01}}} \right\} \\[2ex]
\CMcal{P}_{{\CMcal{B}_2}} &= \chi _{{\CMcal{B}_2}\!{\CMcal{B}_2}} =\! \frac{{{\eta _0}}}{2}\!\!\int\!\!\!\!\!\int\!\! {{{\left| {{\CMcal{B}_2( {\theta{,}\varphi })}} \right|}^2}dS} \! = \!\frac{{{\eta _0}}}{2}\!\!\int\!\!\!\!\!\int \!{{{\left| \boldsymbol{v}_{{\CMcal{B}_2}}^{\{ {{Z_\text{I}},{Z_\text{II}}} \}}\boldsymbol{\CMcal{E}}_\text{emb}^T \right|}^2}\!dS} \nonumber \\
&= \!\frac{{{\eta _0}}}{2}\!\!\int\!\!\!\!\!\int\!\! {{{\left| { \ell _{{\CMcal{B}_2}}^{\{ {{Z_\text{I}},{Z_\text{II}}} \}}\left[ {{\CMcal{E}_1( {\theta{,}\varphi })} - {\CMcal{E}_2( {\theta{,}\varphi })}} \right]} \right|}^2}dS} \nonumber\\
& = {\left| {\ell _{{\CMcal{B}_2}}^{\{ {{Z_\text{I}},{Z_\text{II}}} \}}} \right|^2}\left( {{\CMcal{P}_{{\CMcal{E}_1}}} + {\CMcal{P}_{{\CMcal{E}_2}}} - {\chi _{12}} - {\chi _{21}}} \right) \nonumber\\
& = 2{\left| {\ell _{{\CMcal{B}_2}}^{\{ {{Z_\text{I}},{Z_\text{II}}} \}}} \right|^2}\left( {{\CMcal{P}_{{\CMcal{E}_1}}} - {\chi _{21}}} \right).
\end{align}
\end{subequations}

When the basis reactances $X_\text{I}$ and $X_\text{II}$ satisfy the reactive load condition in (\ref{eq:eq44}), using (\ref{eq:eq25}) it can be shown that
\begin{subequations}\label{eq:eqnew46}
\begin{align}
\ell _{{\CMcal{B}_1}}^{\{ {{Z_\text{I}},{Z_\text{II}}} \}}\!=\!\frac{{\CMcal{S}_{01}( \CMcal{S}_{11} - \CMcal{S}_{21} )}^*}{{1\! - \!{{( {{{\CMcal{S}}_{11}}\! -\! {{\CMcal{S}}_{21}}} )}^*}( {{{\CMcal{S}}_{11}} \!+\! {{\CMcal{S}}_{21}}} )}} \\[2ex]
\left|\ell _{{\CMcal{B}_2}}^{\{ {{Z_\text{I}},{Z_\text{II}}} \}}\right|^2\!=\! \left|\frac{ \CMcal{S}_{01}}{{1\! - \!{{( {{{\CMcal{S}}_{11}}\! -\! {{\CMcal{S}}_{21}}} )}^*}( {{{\CMcal{S}}_{11}} \!+\! {{\CMcal{S}}_{21}}} )}} \right|^2.
\end{align}
\end{subequations}
Therefore, the dependency of $\CMcal{P}_{\CMcal{B}_1}$ and $\CMcal{P}_{\CMcal{B}_2}$ on the basis reactances $X_\text{I}$ and $X_\text{II}$ is removed, i.e.
\begin{subequations}\label{eq:eqnew47}
\begin{align}
\CMcal{P}_{{\CMcal{B}_1}} &= {\CMcal{P}_{{\CMcal{E}_0}}} \!+ \!2{\left| \!\frac{{\CMcal{S}_{01}( \CMcal{S}_{11} - \CMcal{S}_{21} )}^*}{{1\! - \!{{( {{{\CMcal{S}}_{11}}\! -\! {{\CMcal{S}}_{21}}} )}^*}( {{{\CMcal{S}}_{11}} \!+\! {{\CMcal{S}}_{21}}} )}} \right|^2}\!\!\left( {{\CMcal{P}_{{\CMcal{E}_1}}} \!+\! {\chi _{21}}} \right) \nonumber\\
&\quad + 4{\mathop{\rm Re}\nolimits} \left\{ {\!\frac{{\CMcal{S}_{01}( \CMcal{S}_{11} - \CMcal{S}_{21} )}^*}{{1\! - \!{{( {{{\CMcal{S}}_{11}}\! -\! {{\CMcal{S}}_{21}}} )}^*}( {{{\CMcal{S}}_{11}} \!+\! {{\CMcal{S}}_{21}}} )}}{\chi _{01}}} \right\} \\[2ex]
\CMcal{P}_{{\CMcal{B}_2}} &= 2\left|\frac{ \CMcal{S}_{01}}{{1\! - \!{{( {{{\CMcal{S}}_{11}}\! -\! {{\CMcal{S}}_{21}}} )}^*}( {{{\CMcal{S}}_{11}} \!+\! {{\CMcal{S}}_{21}}} )}} \right|^2\left( {{\CMcal{P}_{{\CMcal{E}_1}}} - {\chi _{21}}} \right).
\end{align}
\end{subequations}

On the other hand, if the losses in the metallic and dielectric materials of the antenna are negligible (namely, the three-port radiating structure is lossless), thanks to (\ref{eq:eq11}) and (\ref{eq:eq12}), $\CMcal{P}_{\CMcal{B}_1}$ and $\CMcal{P}_{\CMcal{B}_2}$ can be expressed in terms of only the S-parameters and the basis impedances $Z_\text{I}$ and $Z_\text{II}$ at the control ports, i.e.
\begin{subequations}\label{eq:eq46}
\begin{align}
{\CMcal{P}_{{\CMcal{B}_1}}} \! &= 1\! -\! \sum\limits_{n = 0}^2 {{{\left| {{\CMcal{S}_{n0}}} \right|}^2}} - 4{\mathop{\rm Re}\nolimits} \left\{ {\ell _{{\CMcal{B}_1}}^{\{ {{Z_\text{I}},{Z_\text{II}}} \}}\sum\limits_{n = 0}^2 {\CMcal{S}_{n0}^ * {\CMcal{S}_{n1}}} } \right\} \nonumber \\
&\quad+ \!2{\left| {\ell _{{\CMcal{B}_1}}^{\{ {{Z_\text{I}},{Z_\text{II}}} \}}} \right|^2} \!\!  \left[ {1\!\! -\!\! \sum\limits_{n = 0}^2 {{{\left| {{\CMcal{S}_{n1}}} \right|}^2}} \!\! -\!\! \sum\limits_{n = 0}^2 {\CMcal{S}_{n2}^ * {\CMcal{S}_{n1}}} } \right]\\
{\CMcal{P}_{{\CMcal{B}_2}}}\! &= 2{\left| {\ell _{{\CMcal{B}_2}}^{\{ {{Z_\text{I}},{Z_\text{II}}} \}}} \right|^2} \!\! \left[ {1 \!-\! \sum\limits_{n = 0}^2 {{{\left| {{\CMcal{S}_{n1}}} \right|}^2}} \! +\! \sum\limits_{n = 0}^2 {\CMcal{S}_{n2}^ * {\CMcal{S}_{n1}}} } \right].
\end{align}
\end{subequations}




\ifCLASSOPTIONcaptionsoff
  \newpage
\fi



%
%
%
%

\bibliographystyle{IEEEtran}
\bibliographystyle{IEEEtranTCOM}
\bibliography{IEEEabrv,bibliograph2y}


%

\end{document}